\documentclass[aps,pre,twocolumn,superscriptaddress,10pt]{revtex4-2}
\usepackage[utf8]{inputenc}
\usepackage{color}
\usepackage{amsmath}
\usepackage{amssymb}
\usepackage{xcolor} 
\usepackage{graphicx} % Required for inserting images
\usepackage{esint}
\usepackage{appendix}
\usepackage{comment}
\usepackage{bm}
\usepackage{mathtools}

\usepackage[T1]{fontenc}

\usepackage{hyperref}

\makeatletter
\@ifundefined{textcolor}{}
{%
 \definecolor{BLACK}{gray}{0}
 \definecolor{WHITE}{gray}{1}
 \definecolor{RED}{rgb}{1,0,0}
 \definecolor{GREEN}{rgb}{0,1,0}
 \definecolor{BLUE}{rgb}{0,0,1}
 \definecolor{CYAN}{cmyk}{1,0,0,0}
 \definecolor{MAGENTA}{cmyk}{0,1,0,0}
 \definecolor{YELLOW}{cmyk}{0,0,1,0}
}

\makeatother

\RequirePackage{xcolor} %DIF PREAMBLE
\RequirePackage{soul} %DIF PREAMBLE
\sethlcolor{yellow} %DIF PREAMBLE
\soulregister\cite7 %DIF PREAMBLE
\soulregister\ref7 %DIF PREAMBLE
\soulregister\pageref7 %DIF PREAMBLE
\soulregister\eqref7 %DIF PREAMBLE
\soulregister\emph7 %DIF PREAMBLE
\soulregister\textit7 %DIF PREAMBLE
\soulregister\textbf7 %DIF PREAMBLE

\makeatletter %DIF PREAMBLE
\providecommand{\DIFmathhighlight}[1]{% %DIF PREAMBLE
  \begingroup\setlength{\fboxsep}{0.25pt}% %DIF PREAMBLE
  \colorbox{yellow}{\ensuremath{#1}}% %DIF PREAMBLE
  \endgroup} %DIF PREAMBLE
\DeclareRobustCommand{\DIFaddtex}[1]{% %DIF PREAMBLE
  \ifmmode\DIFmathhighlight{#1}\else\protect\hl{#1}\fi} %DIF PREAMBLE
\DeclareRobustCommand{\DIFadd}[1]{\DIFaddtex{#1}} %DIF PREAMBLE
\providecommand{\DIFdel}[1]{} %DIF PREAMBLE
\providecommand{\DIFdelFL}[1]{} %DIF PREAMBLE
\makeatother %DIF PREAMBLE

\begin{document}

\title{Hierarchical Reconstruction of Time-arrow from Multi-time Correlations}
\author{Yijia Cheng}
\affiliation{Department of Chemical Physics, University of Science and Technology of China, Hefei, Anhui 230026, China}
\author{Ruicheng Bao}
\email{ruicheng@g.ecc.u-tokyo.ac.jp}
\affiliation{Department of Physics, Graduate School of Science, The University of Tokyo, Hongo, Bunkyo-ku, Tokyo 113-0033, Japan}
\author{Zhonghuai Hou}
\email{hzhlj@ustc.edu.cn}
\affiliation{Department of Chemical Physics, University of Science and Technology of China, Hefei, Anhui 230026, China}
\affiliation{Hefei National Research Center for Physical Sciences at the Microscale, University of Science and Technology of China, Hefei, Anhui 230026, China}

\begin{abstract}
The entropy production rate, a key measure of thermodynamic irreversibility in thermodynamics of small systems, is difficult to determine directly in experiments, motivating lower-bound-based estimation from observations.
However, a systematic framework for organizing increasing amounts of the irreversibility information in experimental state observables into progressively tighter bounds remains lacking.
Here, we show that multi-time correlations of a class of state observations naturally encode this information to provide a hierarchy. 
By defining a reconstruction operation as a combination of correlations, we obtain a sequence of lower bounds on the entropy production rate. 
Correlations of higher order capture the thermodynamic information at greater temporal depth, thereby capturing more irreversibility and yielding tighter bounds. 
Under ideal conditions, this hierarchy converges to the full entropy production rate in the limit of infinitely dense observations over a finite time window.
\end{abstract}

\maketitle

\textit{Introduction.}--
Thermodynamic irreversibility, manifested macroscopically as the arrow of time \cite{eddington1927nature}, has a microscopic description through the entropy production rate (EPR) of stochastic trajectories \cite{seifert2005entropy, seifert2012stochastic}. Within thermodynamics of small systems, the entropy production rate is the central quantitative measure of nonequilibrium heat dissipation, with broad relevance from biomolecular processes \cite{sartori2015free, roldan2021quantifying, hathcock2024time, thipmaungprom2025active, cao2025stochastic} and soft matter \cite{tociu2019dissipation, nguyen2021organization, nardini2017entropy, ro2022model, loos2023nonreciprocal, goswami2024anomalous, dieball2025precisely} to heat engines \cite{gingrich2014efficiency, martinez2016brownian, pietzonka2018universal, takaki2022information, tohme2025gambling, liang2025minimal, ito2025universal}. Yet determining the entropy production rate typically requires detailed knowledge of the underlying stochastic dynamics—information that is rarely available experimentally \cite{seifert2005entropy, seifert2012stochastic, seifert2019stochastic}. This has motivated thermodynamic inference: the program of bounding the entropy production rate from experimentally accessible observables without reconstructing the full dynamics \cite{seifert2019stochastic, dieball2025perspective}. This field has advanced rapidly in recent years \cite{dieball2025perspective, dieball2025thermodynamic, godec2022challenges, blom2024milestoning, harunari2022learn, roldan2021quantifying, song2024information}.

A particularly appealing goal within this program is to construct hierarchical bounds, which tighten monotonically as richer information is incorporated \cite{seifert2019stochastic, skinner2021improved}. Such hierarchies have been explored for path observables, such as transition statistics \cite{bisker2017hierarchical, skinner2021improved, igoshin2025uncovering}, waiting-time distributions \cite{skinner2021estimating, nitzan2023universal} and trajectory observables \cite{aguilera2026inferring}.
A small number of results also exist that work directly on sequences of observed states themselves, demonstrating that time-reversal asymmetry can yield lower bounds on the entropy production rate \cite{roldan2010estimating, roldan2021quantifying, kapustin2024utilizing, seifert2025universal}. The missing step, however, is a framework based on this thermodynamic content in terms of experimentally standard state observables, despite their practical importance \cite{seifert2012stochastic, singh2024inferring, di2025force}. The key point is that, in such measurements, irreversibility lies not in a single state alone, but in the time-ordered correlation structures among observables that resolve multiple underlying states. The open question is therefore whether this observable arrow of time can itself be organized into a systematic hierarchy that progressively reconstructs thermodynamic irreversibility.

Here we answer this question by establishing a monotonic hierarchy of lower bounds on the entropy production rate $\dot\sigma$ for nonequilibrium steady states built from multi-time correlation functions of observables, which are accessible at many experimental facilities \cite{qian2004fluorescence, qian2009chemical, ball1988single, yao2009microrheology}. We focus on a broadly applicable class of observables with constant sum \cite{aitchison1982statistical, egozcue2006simplicial}, with the potential for broad applicability \cite{chodera2014markov, ting2008two, song2024information, kispersky2008stochastic, egan2022stochastic, qian2010chemical, szymanska2015effective} and physical relevance \cite{talaga2007markov, song2024information, sanabria2020resolving, bao2025measuring, esposito2012stochastic, gingrich2017inferring, martinez2019inferring, skinner2021improved, blom2024milestoning, dieball2025perspective}. For such observables, we establish that the asymmetry of (n)-time correlations under time reversal \cite{qian2004fluorescence, qian2009chemical, ohga2023thermodynamic, gu2024thermodynamic, bao2025measuring} can define an optimal reconstructed quantity, $\dot\sigma^{\mathrm{est-}n}_\mathrm{opt}$, such that
\begin{equation}
    \dot\sigma^{\mathrm{est-}1}_\mathrm{opt} \le \dot\sigma^{\mathrm{est-}2}_\mathrm{opt} \le \cdots \le \lim_{n \to \infty} \dot\sigma^{\mathrm{est-}n}_\mathrm{opt} \le \dot\sigma ,
    \label{intro}
\end{equation}
where the increasing $n$ corresponds to using more sampling points for optimal estimation.
Under ideal conditions, the final inequality is saturated, so the hierarchy converges to the true $\dot\sigma$ in the limit of complete temporal information. Each additional observation time therefore contributes a definite, quantifiable portion of the irreversibility visible in the measured dynamics, yielding systematically tighter bounds on dissipation.

Beyond providing a practical framework for thermodynamic inference, our results demonstrate that the arrow of time in observed dynamics is not merely a phenomenological signature of nonequilibrium, but a quantitatively thermodynamic one. The temporal asymmetries encoded in measured multi-time correlations reveal how much irreversibility is visible at a given observational level, and thus how much of the true $\dot\sigma$ can be captured. Our hierarchy therefore assigns thermodynamic meaning to progressively more detailed temporal structure in experimental data, linking the observed dynamical evolution direction to a fundamental thermodynamic measure.

\begin{figure*}[!bth]
    \centering
    \includegraphics[width=\textwidth]{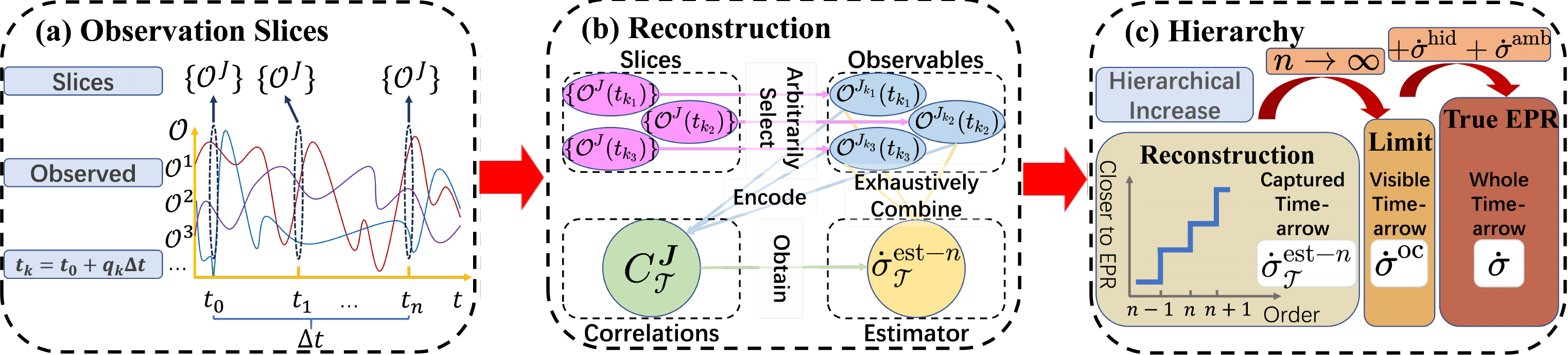}
    \caption{
    (a) From observations to observation slices. The evolving observables are sampled at time points $t_k$, forming observation slices. (b) From observation slices to reconstruction. Multi-time correlations are constructed from selections of each slice, and then we can obtain the reconstruction. (c) Hierarchical reconstruction of the time-arrow. Each additional sampling time point (step) adds temporal information and tightens the reconstruction, bringing it closer to the whole time-arrow. When $n \to \infty$ and $\dot{\sigma}^{\text {hid }} = \dot{\sigma}^{\text {amb }}=0$, the reconstruction reaches the true entropy production rate (EPR).
    }
    \label{Schem}
\end{figure*}

\textit{Setup.}--
Consider a class of unprocessed (``raw'') non-negative state observables with homogeneous units and constant sum, $\left\{
\mathcal{O}^{\mathrm{raw-}1}(t),
\mathcal{O}^{\mathrm{raw-}2}(t),
\ldots,
\mathcal{O}^{\mathrm{raw-}N_J}(t)
\right\}$. 
The duration of these measurements must be considerably shorter than the timescale of microscopic dynamics.
After normalization by this sum,
$\mathcal{O}^{J}(t)
=
\mathcal{O}^{\mathrm{raw-}J}(t) /
\sum_{K}\mathcal{O}^{\mathrm{raw-}K}(t)$,
with physical units removed, the data can be represented in the form
\begin{equation}
\forall J,\quad \mathcal{O}^J(t)\ge 0,\qquad \sum_{J=0}^{N_J}\mathcal{O}^J(t)=1,
\label{observable_condition}
\end{equation}
which holds at ALL times. 
This is a generalizable operation that remains unaffected even by tiny components (but not recommended; see S1 of \cite{supp} for discussion). 
The constant-sum condition selects observables encoding micro-dynamics, and normalization purges irrelevant amplitude to expose fluctuations, which characterize the irreversibility of stochastic systems. The construction of this variable relies on measuring the physical interface of conserved quantities, as exemplified later; it fails when such conservation or measurement lapses.

Such observables arise naturally in many experimental settings, including single-molecule conformational dynamics \cite{chodera2014markov, ting2008two, song2024information, xie2024review}, ion-channel gating kinetics \cite{kispersky2008stochastic, o2014systematic}, and chemical reaction networks \cite{egan2022stochastic, qian2010chemical, szymanska2015effective, chun2023trade}.
In these contexts, the observables are best viewed as shares of a fixed sum: at each time, the signal is redistributed among detection channels \cite{talaga2007markov, song2024information}, subpopulations \cite{sanabria2020resolving, bao2025measuring}, or coarse-grained states \cite{esposito2012stochastic, gingrich2017inferring, martinez2019inferring, blom2024milestoning, yang2024single, dieball2025perspective}. 
For instance, we can define the spatial particle-number distribution of a closed space \cite{bao2025measuring} as the observables here, based on particle number conservation.
Since the data live on a simplex that describes compositions of a whole, they can be called composition-like \cite{aitchison1982statistical, egozcue2006simplicial} in their data structure. This makes composition-like observables a natural setting for thermodynamic inference based on state observables, since the time-arrow is more likely to emerge.
This allows observations to be recast in the language of information thermodynamics and eliminate the risk of null results.
Under additional restrictions, our framework can be extended to cases in which the sum varies in time, ensuring its wide practicality; see S1 of \cite{supp} for details.

We regard all measured observables $\mathcal{O}^{J_k}(t)$ ($J_k \in \left \{1,2,\cdots, N_J\right \}$) recorded at one sampling time as an observation slice (indexed by $k$, with a total number of $n+1$). Choose a window $[t,t+\Delta t]$ and relative fractions 
$0=q_0\le q_1\le\cdots\le q_n=1$. 
Writing $\mathcal T=(t+q_0\Delta t,\ldots,t+q_n\Delta t)$ for this deterministic schedule and $\bm J=(J_0,\ldots,J_n)$ for a sequence of observations, we then build $n$-th order temporal correlations by selecting one entry from each of several slices [Fig. \ref{Schem}-(a)~(b)]
\begin{equation}
    C^{\bm J}_{\mathcal T} \coloneqq \left \langle\prod_{k=0}^n\mathcal{O} ^{J_k}(t+q_k\Delta t)\right \rangle,
    \label{average}
\end{equation}
uniformly handling auto- and cross-correlation functions.

\begin{figure}[!bth]
    \centering
    \includegraphics[width=1\columnwidth, keepaspectratio]{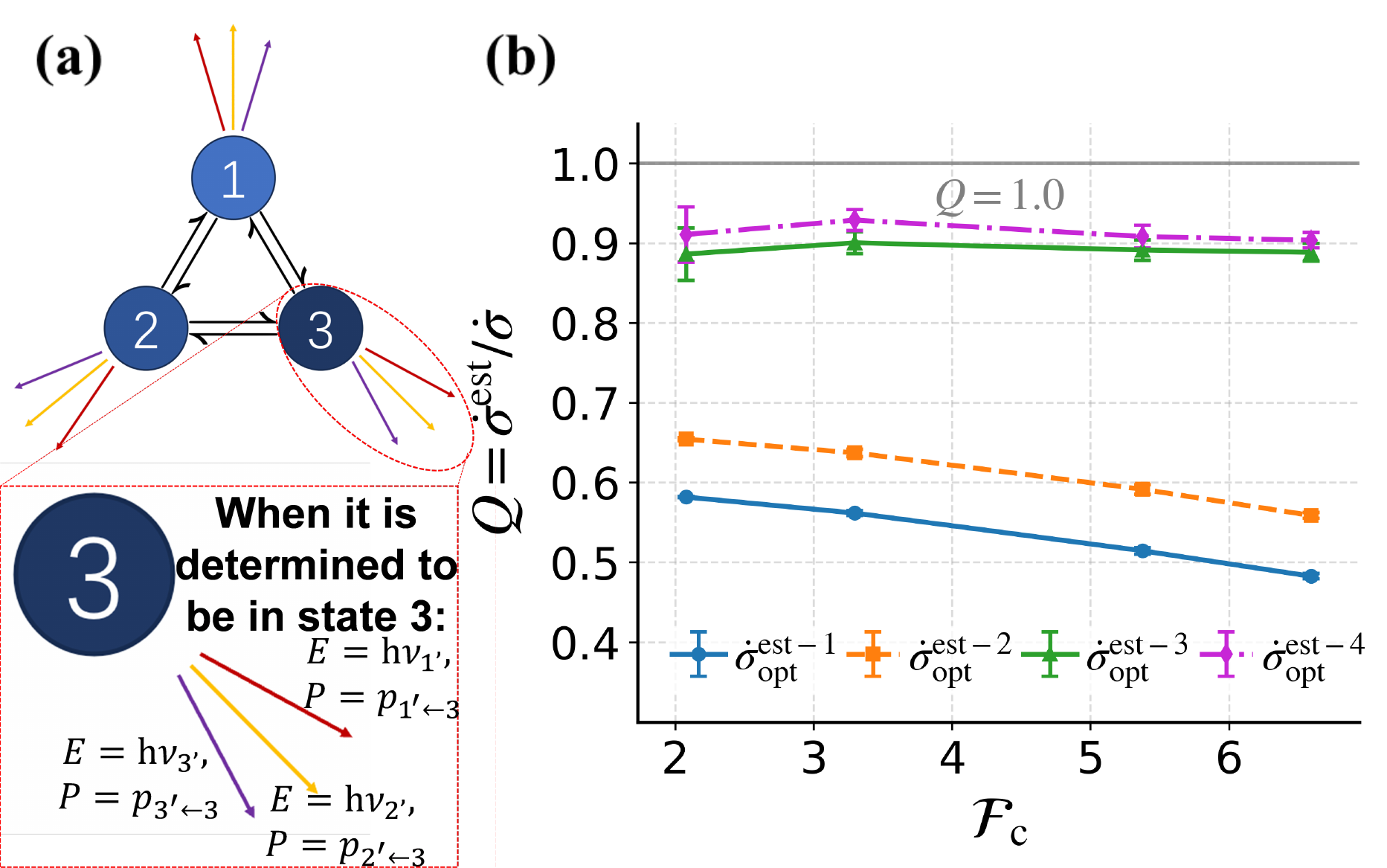}
    \caption{(a) Biomolecular process with many-to-many state-signal correspondence. Each microstate (circles in shades of blue) can emit photons detected in different color channels (colored arrows), with probabilities $p_{J\gets i}^\mathrm{map}$ satisfying $\sum_Jp_{J\gets i}^\mathrm{map} =1$. (b) Optimized reconstruction for different orders $n$ under different cycle affinities. All reconstructed values remain below the true entropy production rate (EPR), and increase with $n$, approaching $\dot \sigma$. Error bars indicate the standard deviation over five independent simulations. Note that: 1) In steady states, $\dot \sigma^\mathrm{est}(t)$ and $\dot \sigma(t)$ are time-independent, so ``$(t)$'' is omitted; 2) Though error bars of $\dot \sigma^\mathrm{est-3}_\mathrm{opt}$ and $\dot \sigma^\mathrm{est-4}_\mathrm{opt}$ overlap, $\dot \sigma^\mathrm{est-4}_\mathrm{opt} > \dot \sigma^\mathrm{est-3}_\mathrm{opt}$ holds for every single trajectory.}
    \label{eg}
\end{figure}

We model each observable as an instantaneous-state-dependent but time-independent readout of the underlying microstate. Specifically, measurements probe the system and collect information via interfaces that map subsets of microstates onto signals. We refer to these information-transmitting pathways as ``observation channels'' \cite{talaga2007markov}, with the fluorescent colors in Fig. \ref{eg} serving as a visual example. Formally, each microstate $i$ carries a fixed observation profile $\mathcal{O}^J_i$, which specifies how that state distributes weight across the observation channels:
\begin{equation}
        \forall i \in \left\{ 1, 2, \cdots, N_i \right\}, \qquad \mathcal{O}^J_i \ge 0 , \qquad \sum_J\mathcal{O}^J_i = 1.
    \label{state_condition}
\end{equation}
Denoting $\bm i=(i_n,\ldots,i_1,i_0)$, we can then write the resulting multi-time correlation as
\begin{equation}
    \begin{split}
        &C^{\bm J}_{\mathcal T} = \sum_{\bm i} \mathcal{O}^{J_n}_{i_n} \ldots \mathcal{O}^{J_1}_{i_1} \mathcal{O}^{J_0}_{i_0}\\
        &\times\mathcal{P}(i_n,t+q_n\Delta t;\ldots;i_1,t+q_1\Delta t;i_0,t+q_0\Delta t).
    \end{split}
    \label{correlation-raw}
\end{equation}
Here, $\mathcal{P}(i_n,t+q_n\Delta t;\ldots;i_1,t+q_1\Delta t;i_0,t+q_0\Delta t)$ denotes the joint probability of the microstate; given $\mathcal T$ as fixed design parameters, each $t+q_k\Delta t$ is the deterministic time of $i_k$, no longer a random variable.

\textit{Stochastic Observation Channel.}--
To make the meaning of the correlation $C^{\bm J}_{\mathcal T}$ more transparent, we reinterpret each observable label $J$ as a stochastic observation channel: when the system is in microstate $i$, the measurement reports channel $J$ with ``mapping'' probability $p_{J \gets i}^\mathrm{map}  \coloneqq \mathcal{O}^J_i$. Then we obtain
\begin{equation}
    p_J^\mathrm{oc}(t) = \sum_i p_{J \gets i}^\mathrm{map} \cdot p_i(t).
    \label{random-mapping}
\end{equation}
Here, $p_J^\mathrm{oc}(t)$ is the probability of reporting channel $J$ (with the superscript ``oc'' meaning observation channel), yielding a readout of $J$ when observation timescale is much longer than that of microscopic dynamics. 

This many-to-many mapping from states to observables has been used previously by Bao et al. for many-body interacting systems \cite{bao2025measuring} and by Song et al. for fluorescent signals of molecular machines \cite{song2024information}, even in the hidden Markov model of speech recognition \cite{rabiner2002tutorial, yang2025transformer, wang2025normalization}. None of the works, however, constructs the mapping bottom-up, valid under a broad class of state observables.

According to Eq. \eqref{random-mapping} and Eq. \eqref{correlation-raw}, we find that the correlation corresponds to the probability of a specified sequence of observation channels (given fixed sampling times $\mathcal T$) $(J_n,t+q_n\Delta t;...;J_1,t+q_1\Delta t;J_0,t+q_0\Delta t)$, i.e. the occurrence probability of the event $\omega^{\bm J}_{\mathcal T}$:
\begin{equation}
    C^{\bm J}_{\mathcal T} = P (\omega^{\bm J}_{\mathcal T}).
    \label{probability-meaning}
\end{equation}
That is, a multi-time correlation is not just an abstract moment: it is the probability of seeing a particular sequence of channels at the chosen sampling time points, constituting a further reduction of information contained in partially observed channel trajectories. It can be considered as a special extension of trajectory coarse-graining \cite{seifert2025universal}, which is practically unified with the correlation functions of observables measured in experiments.

\textit{Time-arrow Reconstruction: Lower Bounds of EPR.}--
In stochastic thermodynamics, setting $\mathrm{k_B}=1$, the entropy production of Markov jump dynamics is given by $\sigma_{\left [ 0,t \right ] } = D_{\mathrm{KL}}(\mathbb{P}_\gamma\|\mathbb{P}^\dagger_\gamma)$, where $D_{\mathrm{KL}}(.\|.)$ denotes the Kullback-Leibler divergence, an information-theoretic measure of the disparity between two probability distributions \cite{polyanskiy2025information}.
%Here, the path integral $\int_{\{\gamma_{\left [ 0,t \right ] }\}} \mathrm{d}\gamma (.)$ is taken over the space of Markovian trajectories $\{\gamma_{\left [ 0,t \right ] }\}$, whose elements are determined by underlying Markovian jump dynamics; and $D_{\mathrm{KL}}(.\|.)$ denotes the Kullback–Leibler divergence. 
It measures how distinguishable forward trajectories are from their time-reversed counterparts, thereby characterizing the time-arrow.
Entropy production rate, naturally, is defined as its time derivative $\dot\sigma(t) = \mathrm{d} \sigma_{\left [ 0,t \right ] }/\mathrm{d} t$.

We therefore build an experimentally accessible estimator of the entropy production rate from multi-time correlations, calling it time-arrow reconstruction. For a chosen set of sampling time points within a window $\Delta t$, with $q_k^\leftarrow=1-q_{n-k}$, $\mathcal T^\leftarrow=(t+q_0^\leftarrow\Delta t,\ldots,t+q_n^\leftarrow\Delta t)$, 
$J_k^\leftarrow=J_{n-k}$ and $\bm J^\leftarrow=(J^\leftarrow_0,\ldots,J^\leftarrow_n)$, the estimator compares the correlations with those of its time-reversed counterpart, which can be written in a unified form as [Fig. \ref{Schem}-(b)]
\begin{equation}
    \dot\sigma^{\mathrm{est-}n}_{\mathcal T}(t) \coloneqq \sum_{\bm J} \frac{C^{\bm J}_{\mathcal T}}{\Delta t} \ln \frac{C^{\bm J}_{\mathcal T}}{C^{\bm J^\leftarrow}_{\mathcal T^\leftarrow}},
    \label{estimator}
\end{equation}
where the superscript ``est-$n$'' indicates the reconstruction based on the slice number $n+1$. This is the DEFINING equation in this study, wherein the notation in Eq. \eqref{intro} is established. Its time integral can serve as a representation of time-arrow signature during $\Delta t$ in steady states.

Here, we establish how it reconstructs the entropy production rate. Since the event $\omega^{\bm J}_{\mathcal T}$ is obtained by mapping and sampling from the trajectory $\gamma$, and $\dot\sigma(t)$ is time-independent in steady states, the data-processing inequality \cite{polyanskiy2025information} gives (see S2 of \cite{supp})
\begin{equation}
    \dot\sigma(t) \ge \dot\sigma^{\mathrm{est-}n}_{\mathcal T}(t).
    \label{bounds}
\end{equation}
This is our first main result: for any choice of $n$, $\Delta t$ and $\{q_k\}_{k=0}^n$ (i.e., arbitrary sampling time setting $\mathcal T$), the reconstruction provides a lower bound on the true entropy production rate.
Any reduced or coarse-grained observation blurs part of the forward–backward asymmetry in the underlying trajectories, so the inferred irreversibility can only decrease.
In this way, we establish a practical tool for thermodynamic inference: one can measure the correlations, compute $\dot\sigma^{\mathrm{est-}n}_{\mathcal T}(t)$, and thereby estimate the approximate range of $\dot\sigma(t)$.

\textit{Hierarchical Reconstruction of Time-arrow.}--
As the number of slices increases, more resolvable observation sequences can be obtained, therefore retaining more of the temporal asymmetry.
A quantitative description follows, which establishes the hierarchy in Eq. \eqref{intro}:
\begin{equation}
    \sup_\mathcal T\{\dot\sigma^{\mathrm{est-}n+1}_{\mathcal T}(t)\} \ge \sup_{\mathcal T}\{\dot\sigma^{\mathrm{est-}n}_{\mathcal T}(t)\},
    \label{sup}
\end{equation}
where $\sup_{\mathcal T}\{\dot\sigma^{\mathrm{est-}n}_{\mathcal T}(t)\}$ is exactly $\dot\sigma^{\mathrm{est-}n}_\mathrm{opt}$ of Eq. \eqref{intro}, being the result of optimizing $\Delta t$ and $\{q_k\}_{k=0}^{n}$.
This inequality is the second main result of the article. 
According to Eq. \eqref{probability-meaning}, with more sampling time points, the observed sequence constrains the hidden microscopic trajectory more tightly, so the resulting estimator systematically moves closer to the true entropy production rate [Fig. \ref{Schem}-(c)]. 
It guarantees the monotonic refinement of the reconstruction. We further demonstrate in the End Matter how optimization over $\Delta t$ and $\{q_k\}$ works.

In the limit of infinitely many observation slices over a finite interval $\Delta t$, the sampled observation sequence becomes an entire observation-channel trajectory. The reconstruction then compares the whole ensembles of the forward and reversed trajectories.
According to Eq. \eqref{sup}, we further obtain
\begin{equation}
    \dot \sigma^\mathrm{oc}_{\Delta t}(t) = \lim_{\substack{n \to \infty \\ \sup r_k \to 0}}\dot\sigma^{\mathrm{est-}n}_{\mathcal T}(t) \ge \dot\sigma^{\mathrm{est-}n}_{\mathcal T}(t),
    \label{limit}
\end{equation}
where $r_k \coloneqq q_k-q_{k-1}$ represent the relative distance of sampling times. $\dot \sigma^\mathrm{oc}_{\Delta t}(t)$ is the whole reconstruction of time arrow [Fig. \ref{Schem}-(c)]. The definition of $\dot \sigma^\mathrm{oc}_{\Delta t}(t)$, is determined by the Kullback-Leibler divergence of stochastic channel trajectories (see S4.1 of \cite{supp} for details).

Furthermore, for a stationary finite-state Markov jump process, we prove analytically (S4.4 of \cite{supp}) that the following single condition SUFFICES for
\begin{equation}
    \dot{\sigma}^{\mathrm{oc}}_{\Delta t}=\dot{\sigma}:
    \label{oc-limit}
\end{equation}
for all $i_1\neq i_2$, there exists $J$ such that $\mathcal{O}^{J}_{i_1}\neq\mathcal{O}^{J}_{i_2}$, i.e., no two microstates share the same observation profile.
This condition requires neither a one-to-one correspondence nor any exclusive channel (see S4.4 of \cite{supp}, with the schematic figure). 
This implies that repeated readouts in the dense-sampling limit will eliminate the ambiguity. 
In other words, ambiguity in the state–-channel mapping is not a fundamental loss of irreversibility but an amplified loss caused by finite resolution, which can be recovered at the cost of denser sampling.
It reveals the following essential principle of observation: while the system stays in a state $i$, the readouts are independent samples drawn from the same profile; under dense sampling, this dwell contains arbitrarily many readouts, i.e., the channel statistics converge to the profile, thereby identifying $i$ (see S4.4 for strict derivation). 
However, states with identical profiles cannot be distinguished at any resolution: they constitute an unrecoverable coarse-graining, for which $\dot\sigma^{\mathrm{oc}}_{\Delta t}$ can remain below $\dot\sigma$.

\textit{How Far from Whole EPR.}--
Within our framework, the gap between a finite-slice reconstruction and the whole entropy production rate has only two sources: incomplete temporal sampling and ambiguity in the state-to-observation mapping.
Since Eq. \eqref{limit} fully accounts for the former source, the entropy production rate can then be decomposed as \cite{bao2025measuring}
\begin{equation} 
\dot \sigma(t) = \dot \sigma^\mathrm{oc}_{\Delta t}(t) + \dot \sigma^\mathrm{hid}_{\Delta t}(t) + \dot \sigma^\mathrm{amb}_{\Delta t}(t), 
\end{equation}
where $\dot \sigma^\mathrm{hid}_{\Delta t}(t)$ denotes the irreversibility ``hidden'' inside each observation channel, and $\dot \sigma^\mathrm{amb}_{\Delta t}(t)$ represents the irreversibility hidden between transition paths that are undistinguishable during observation, as the part of invisible irreversibility caused by ``ambiguity'' \cite{song2024information} (see S4.1 of \cite{supp} for detailed definition). Both terms are non-negative, and the violation of Eq. \eqref{oc-limit} (discrepancy between $\dot \sigma^\mathrm{oc}(t)$ and the true entropy production rate caused by identical-profile states) is entirely due to these two layers of the underestimation [Fig. \ref{Schem}-(c)].

When each state corresponds to a unique observation channel (i.e., in the deterministic lumping coarse-graining framework of stochastic thermodynamics \cite{esposito2012stochastic, seifert2019stochastic, martinez2019inferring}), $\dot \sigma^\mathrm{amb}(t)$ generally remains nonzero, because the ambiguity of transition paths between two channels may still exist. It vanishes when each pair of channels is connected by at most one microscopic transition. Furthermore, when the microstates can be observed directly \cite{roldan2010estimating, roldan2012entropy}, $\dot \sigma^\mathrm{hid}(t)=\dot \sigma^\mathrm{amb}(t)=0$, and the fully observed time-arrow $\dot \sigma^\mathrm{oc}(t)$ reconstructs the whole $\dot \sigma(t)$ without any hidden contribution. This can be considered as the special case of Eq. \eqref{oc-limit}'s condition, as the correspondence of channels and states is one-to-one.

\textit{An Illustrative Example.}--
In fluorescence measurements, photons from a molecular state are not always in its ``own'' color channel due to the overlap phenomenon \cite{valm2016multiplexed, gunewardene2011superresolution} and experimental effects (e.g., cross-talk \cite{hellenkamp2018precision}). The state–signal correspondence is therefore naturally many-to-many. To illustrate our framework, we adopt the continuous model based on Ref. \cite{song2024information} here.

We consider a minimal three-state biomolecular process governed by master equations. Each microstate has a small probability that its photons are recorded in the ``wrong'' channels. This imperfect color assignment is described by a ``recoloring matrix''
\begin{equation}
    R = \begin{pmatrix} p_{1'\gets 1}^\mathrm{map} & p_{1'\gets 2}^\mathrm{map} & p_{1'\gets 3}^\mathrm{map}\\ p_{2'\gets 1}^\mathrm{map} & p_{2'\gets 2}^\mathrm{map} & p_{2'\gets 3}^\mathrm{map}\\ p_{3'\gets 1}^\mathrm{map} & p_{3'\gets 2}^\mathrm{map} & p_{3'\gets 3}^\mathrm{map}\end{pmatrix},
\end{equation}
where the diagonal entries ($\simeq 1$) give the correct assignment probability, whereas off-diagonals quantify rare erroneous assignments. The experimentally accessible observables are then the fluorescence intensities in the three color channels, whose multi-time correlations may be measured by single-molecule fluorescence correlation spectroscopy [Fig. \ref{eg}-(a)] \cite{qian2004fluorescence, qian2009chemical}. Further details are shown in S6 of \cite{supp}.

Although Eqs. \eqref{limit}-\eqref{oc-limit} imply all the estimators in this example eventually converge to $\dot{\sigma}$ in the dense-observation limit, our numerical results in S6.3 of \cite{supp} suggest even a small recoloring probability may substantially slow their convergence to $\dot{\sigma}$.
We therefore choose a near-ideal recoloring matrix $R_0$ (weak ambiguity) to represent an experimental setting.  
Specifically, we simulate the stochastic trajectories using the Gillespie algorithm \cite{gillespie1977exact} under different cycle affinities $\mathcal{F}_\mathrm{c}$ \cite{schnakenberg1976network,seifert2012stochastic} (adjusted by a one-parameter protocol that is discussed in S6.2 of \cite{supp}), compute the multi-time correlation functions, and thereby obtain both the steady-state entropy production rate $\dot\sigma(t)$ and the optimized reconstructions $\dot \sigma^{\mathrm{est-}n}_{\mathcal T-\mathrm{opt}} (t)$.
For each order $n$, we use the $\mathcal T_\mathrm{opt}$ (see End Matter for optimization method, and S5 of \cite{supp} for details).

As shown in Fig. \ref{eg}-(b), for various $\mathcal{F}\mathrm{c}$, the reconstructed values remain below the true entropy production rate, as required by Eq. \eqref{bounds}. Crucially, for fixed $\mathcal{F}\mathrm{c}$, the reconstruction increases systematically with $n$, in accordance with Eq. \eqref{sup}. Already at $n=4$, the reconstruction captures more than 90\% of the true entropy production rate, i.e., $Q=\dot{\sigma}^{\mathrm{est}\text{-}n}_{\mathrm{opt}}(t)/\dot{\sigma}(t)>0.9$, while the additional gain from $n=3$ to $n=4$ is marginal. This shows that a modest number of observation slices is sufficient to recover most of the experimentally accessible time-arrow information and enables effective inference.

\textit{Advantages of the Bounds.}--
It is illuminating to place our framework in the context of established thermodynamic bounds. A major class of such bounds arises from rearranging thermodynamic uncertainty relations \cite{barato2015thermodynamic, gingrich2016dissipation, koyuk2019operationally, van2020entropy, horowitz2020thermodynamic}. In particular, Shiraishi proved that the pseudo-EPR $\dot\sigma^\mathrm{pseudo}(t)$ provides the optimal inequality; i.e. as a bound on the true entropy production rate, $\dot\sigma^\mathrm{pseudo}(t)$ is tighter than all others through thermodynamic uncertainty relations \cite{shiraishi2021optimal, shiraishi2023introduction}. Moreover, $\dot\sigma^\mathrm{pseudo}(t)$ also appears as an intermediate term in Gu’s inequality for the asymmetry of correlation functions \cite{gu2024thermodynamic}.

From this perspective, it is a key feature that our framework can yield bounds even tighter than $\dot\sigma^\mathrm{pseudo}(t)$. In particular, when the state--observable correspondence is ONE-TO-ONE, i.e. $\exists\, f\ \text{s.t.}\ O_i^J=\delta_{J,f(i)}$, we obtain
\begin{equation}
\begin{split}
\lim_{\Delta t \to 0} \dot\sigma^{\mathrm{est-}n}_{\mathcal T}(t) \ge \dot\sigma^\mathrm{pseudo}(t),
\label{pseudo}
\end{split}
\end{equation}
where $f$ is a bijection from states to observables (see S7.2 of \cite{supp} for the proof).
Even when this correspondence deviates from one-to-one slightly, our bound may still retain its advantage, since it need not overturn the inequality [see Fig. S3]. Even if it does, the advantage is not necessarily lost, because experimental ambiguity (not one-to-one correspondence) also blurs the observed current and loosens the inequalities, thereby driving these inference away from the pseudo-EPR limit \cite{gingrich2016dissipation, shiraishi2021optimal}. This advantage is especially pronounced far from equilibrium, showing that our approach is not merely a reformulation of existing thermodynamic constraints, but can extract more information from the asymmetry of correlations and genuinely sharpen thermodynamic inference beyond known bounds. (See S7 of \cite{supp} for detailed explanation.)

\textit{Discussion and Outlook.}--
In this work, we developed a hierarchical approach to reconstruct the entropy production rate from partially observed state observables through multi-time correlation functions. 
This construction yields a sequence of monotonic lower bounds that systematically tighten as more observations are included. 
In this sense, the inferred bound represents the reconstructible part of the entropy production rate captured by the observation record. 
In the dense-observation limit, the whole entropy production rate is recovered under ideal conditions.
We also show that our approach offers advantages over several existing inference methods.
These results establish that the observed time-reversal symmetry breaking is not merely a signature or emergence of the arrow of time, but also a quantitatively accessible imprint of nonequilibrium dissipation. 
Beyond thermodynamic inference, this work highlights the profound informational depth of multi-time correlation functions, rooted in a long-standing tradition of distilling fundamental insights from correlations \cite{caldeira1983quantum, maldacena2016bound, noe2008probability}.

A number of open problems remain. 
The present formulation is derived for steady states and should be extended to nonstationary or time-dependent settings. 
It is currently also limited to Markov jumps, as well as overdamped Langevin dynamics for some of the results, leaving underdamped Langevin and non-Markov dynamics as important directions for future work. 
In addition, while higher-order correlations improve reconstruction accuracy, they also increase statistical (computational) and experimental cost, calling for more robust inference schemes or accuracy--cost trade-off strategies. 
More broadly, it will be worthwhile to generalize the theory to other varieties of observation.

\textit{Acknowledgements.}--Y.C. and Z.H. are supported by NSFC(22533005). The simulations were performed on the robotic AI-Scientist platform of Chinese Academy of Sciences.
R.B. is supported by JSPS KAKENHI Grant No. 25KJ0766.
We thank Zhiyu Cao for inspiring discussions. 
We thank Mingrui Zhao for helpful pointers on the manuscript. 
We thank Jie Gu for suggestions on enhancing the rigor of the framework.
Y.C. thanks Yi Yang and Caisheng Cheng for insightful guidance on the figure design, and Qin Xu for inspiring comments.
We also acknowledge ChatGPT (5.4, 5.6-Sol, 6-Astra) and Gemini (3 Pro, 3.1 Pro) for derivation checking and code debugging; ChatGPT and Claude (Opus 5.5) for inspiration of narrative/plotting and proofreading; and Gemini and DeepSeek (V4-Pro) for guidance on language polishing.

\bibliography{refs}

\appendix

\section*{End Matter}
This part aims to demonstrate why one must consider larger $n$ when performing thermodynamic inference, despite the accompanying experimental challenges.
The part also shows how to sharpen Eq. \eqref{n+} to Eq. \eqref{sup} beyond the data processing inequality (or other isomorphic methods \cite{roldan2010estimating, roldan2012entropy, roldan2021quantifying, kapustin2024utilizing}) for the interested reader.

\textit{Optimization of Reconstruction}--
To show the temporal details of the plug-in points, we unfold $\mathcal T$, rewriting the $\dot\sigma^{\mathrm{est-}n}_{\mathcal T}(t)$ in the main text as $\dot\sigma^{\mathrm{est-}n}_{\Delta t, \{q_k\}_{k=0}^n}(t)$ (i.e., $\mathcal T \equiv  \mathcal T(\Delta t,\{q_k\}_{k=0}^n)$ and $\dot\sigma^{\mathrm{est-}n}_{\mathcal T}(t)\equiv\dot\sigma^{\mathrm{est-}n}_{\Delta t, \{q_k\}_{k=0}^n}(t)$).

While keeping $\Delta t$ fixed, by adding an intermediate measurement time $t+q^\prime_{n+1}\Delta t$ with $0\le q^\prime_{n+1} \le 1$, one obtains
\begin{equation}
    \dot\sigma^{\mathrm{est-}n+1}_{\Delta t, \{q_k\}_{k=0}^n\cup \{q^\prime_{n+1}\}}(t) \ge \dot\sigma^{\mathrm{est-}n}_{\Delta t, \{q_k\}_{k=0}^n}(t).
    \label{n+}
\end{equation}
The full derivation is in S3 of \cite{supp}.
It implies that tuning $\Delta t$ and $\{q_k\}$ may also tighten the lower bound; however, increasing $n$ plays an irreplaceable role in systematically driving the reconstruction $\dot\sigma^{\mathrm{est-}n}_{\Delta t, \{q_k\}_{k=0}^n}(t)$ closer to the true $\dot{\sigma}(t)$. We can obtain this result via the following steps.

First, for fixed $n$, 
S5.1 and S5.2 of \cite{supp} establish under the stated regularity assumptions that
\begin{subequations}
\begin{align}
    \dot \sigma^{\mathrm{est-}n}_{\Delta t, \{q_k\}_{k=0}^n} (t) \ge 0,\label{optimal_Delta_t-1}\\
    \lim_{\Delta t \to \infty} \dot\sigma^{\mathrm{est-}n}_{\Delta t, \{q_k\}_{k=0}^n}(t) =0,\label{optimal_Delta_t-2}\\
    \lim_{\Delta t \to 0} \dot\sigma^{\mathrm{est-}n}_{\Delta t, \{q_k\}_{k=0}^n}(t) < \infty .\label{optimal_Delta_t-3}
\end{align}
\end{subequations}
Under the assumptions of S5.2 of \cite{supp}, the maximum is attained on the extended domain $\Delta t\ge0$, with the value at $\Delta t=0$ defined by the limit $\Delta t\to0^+$.
Consequently, the optimal parameters $\Delta t_{\mathrm{opt}-n}$ and $(\{q_k\}_{k=0}^n)_{\mathrm{opt}}$ are well-defined as the values at which the reconstruction attains its maximum. 
In the case of complete datasets under sufficiently controlled experimental conditions, these optima can be approximated numerically via an exhaustive search over the parameter space. If $\Delta t_{\mathrm{opt}-n}=0$, sufficiently small positive windows
approximate the optimal value.

Owing to Eq. \eqref{n+} and the optimization procedure with respect to the parameters $\Delta t$ and $\{q_k\}_{k=0}^n$, we obtain
\begin{equation}
\begin{split}
    &\dot \sigma^{\mathrm{est-}n}_{\Delta t_{\mathrm{opt}-n}, (\{q_k\}_{k=0}^n)_{\mathrm{opt}}} (t) \\
    \le & \dot \sigma^{\mathrm{est-}n+1}_{\Delta t_{\mathrm{opt}-n}, (\{q_k\}_{k=0}^n)_{\mathrm{opt}}\cup\{q^\prime_{n+1}\}} (t) \\
    \le & \dot \sigma^{\mathrm{est-}n+1}_{\Delta t_{\mathrm{opt}-n+1}, (\{q_k\}_{k=0}^{n+1})_{\mathrm{opt}}} (t).
\end{split}
\end{equation}
By the definition of optimization, $\dot \sigma^{\mathrm{est-}n}_{\Delta t_{\mathrm{opt}-n}, (\{q_k\}_{k=0}^n)_{\mathrm{opt}}} (t) = \sup_{\Delta t, \{q_k\}_{k=0}^{n}}\{\dot\sigma^{\mathrm{est-}n}_{\Delta t, \{q_k\}_{k=0}^{n}}(t)\}$, which directly yields Eq. \eqref{sup}.
Therefore, upon optimizing $\Delta t$ and $\{q_k\}_{k=0}^n$ at each order $n$, the optimized lower bound remains monotonically tighter with increasing $n$. This emphasizes that a greater number of observation slices enables increasingly efficient thermodynamic inference at its optimum, thereby underscoring the distinct advantage of employing higher $n$ within this framework.

\end{document}

\title{Supplemental Material for ``Hierarchical Reconstruction of Time-arrow from Multi-time Correlations''}
\author{Yijia Cheng}
\affiliation{Department of Chemical Physics, University of Science and Technology of China, Hefei, Anhui 230026, China}
\author{Ruicheng Bao}
\email{ruicheng@g.ecc.u-tokyo.ac.jp}
\affiliation{Department of Physics, Graduate School of Science, The University of Tokyo, Hongo, Bunkyo-ku, Tokyo 113-0033, Japan}
\author{Zhonghuai Hou}
\email{hzhlj@ustc.edu.cn}
\affiliation{Department of Chemical Physics, University of Science and Technology of China, Hefei, Anhui 230026, China}
\affiliation{Hefei National Research Center for Physical Sciences at the Microscale, University of Science and Technology of China, Hefei, Anhui 230026, China}

\maketitle

\setcounter{equation}{0}
\setcounter{figure}{0}
\setcounter{table}{0}
\setcounter{section}{0}

\renewcommand{\theequation}{S\arabic{equation}}
\renewcommand{\thefigure}{S\arabic{figure}}
\renewcommand{\thesection}{S\arabic{section}}
\renewcommand{\thesubsection}{S\arabic{section}.\arabic{subsection}}

\section*{Unfolding the notations}
To expose details and work through the deviation (especially when plugging in time point or constructing conditional probabilities), we unfold the $\bm J$, $\bm i$ and $\mathcal T$. The results of unfolding the notations referenced in the Supplemental Material are presented in the table below.

\begin{table}[h!]
    \centering
    \begin{tabular}{|c|c|c|c|c|}
    \hline
      main text & $C^{\bm J}_{\mathcal T}$ & $C^{\bm J^\leftarrow}_{\mathcal T^\leftarrow}$ & $\omega^{\bm J}_{\mathcal T}$ & $\dot\sigma^{\mathrm{est-}n}_{\mathcal T}(t)$ \\
    \hline
      supplemental material & $C^{J_n,...,J_1,J_0}_{t,\Delta t,\{q_k\}_{k=0}^n}$ & $C^{J_0,J_1...,J_n}_{t,\Delta t,\{1-q_{n-k}\}_{k=0}^n}$ & $\omega^{J_n,...,J_1,J_0}_{t,\Delta t,\{q_k\}_{k=0}^n}$ & $\dot\sigma^{\mathrm{est-}n}_{\Delta t, \{q_k\}_{k=0}^n}(t)$ \\
    \hline
    \end{tabular}
    \caption*{Cross-Reference of Symbols (Main Text and Supplemental Material).}
    \label{tab:placeholder}
\end{table}

\section{Non-negativity and Normalization of $\mathcal{O}^J_i$}
Without loss of generality, we consider an N-state Markov jump process governed by the master equation
\begin{equation}
    \frac{\mathrm{d} \bm{p}(t)}{\mathrm{d} t}  = K\bm{p}(t),
    \label{master}
\end{equation}
where $\bm{p}(t) = \left ( p_1(t), p_2(t),\dots,p_{N_i}(t) \right ) ^\mathrm{T} $ denotes the vector of state probabilities and $K$ is the transition rate matrix. 
The DPI-based bounds and hierarchy also apply to stationary overdamped diffusions with finite forward--reverse path KL divergence, as it represents the continuous limit of the master equation.

When $\{\mathcal{O}^{\mathrm{raw-}J}(t)\}$ are state observables with constant sum, as discussed in the main text, one can obtain
\begin{equation}
    \mathcal{O}^{\mathrm{raw-}J}(t) = \sum_{i=0}^{N_i}\mathcal{O}^{\mathrm{raw-}J}_{i}\delta_{i,i(t)} \ge 0
\end{equation}
and
\begin{equation}
    \sum_{J=0}^{N_J}\mathcal{O}^{\mathrm{raw-}J}(t) = \sum_{i=0}^{N_i}\sum_{J=0}^{N_J}\mathcal{O}^{\mathrm{raw-}J}_{i}\delta_{i,i(t)} = \mathrm{Constant},
\end{equation}
where $\{\mathcal{O}^{\mathrm{raw-}J}_{i}\}$ are time-independent, raw readouts of $J$ in microscopic state $i$. Their duration of measurements must be considerably shorter than the timescale of microscopic dynamics, guaranteeing that a single measurement does not blend different microscopic configurations from separate instants. Since this holds for ALL times, one obtains
\begin{equation}
    \forall i, \qquad \mathcal{O}^{\mathrm{raw-}J}_{i} \ge 0, \qquad \sum_{J=0}^{N_J}\mathcal{O}^{\mathrm{raw-}J}_{i} = \mathrm{Constant}.
\end{equation}
After normalization, one obtains
\begin{equation}
    \mathcal{O}^{J}(t)=\mathcal{O}^{\mathrm{raw-}J}(t)/\sum_{K}\mathcal{O}^{\mathrm{raw-}K}(t)
\end{equation}
i.e.
\begin{equation}
    \mathcal{O}^{J}_i = \mathcal{O}^{\mathrm{raw-}J}_i/\sum_{K}\mathcal{O}^{\mathrm{raw-}K}_i.
\end{equation}
As a result, we show that $\{\mathcal{O}^{J}_i\}$ are also time-independent via 
\begin{equation}
    \sum_{J=0}^{N_J} \mathcal{O}^J_i = \sum_{J=0}^{N_J} \frac{\mathcal{O}^{\mathrm{raw-}J}_i}{\sum_{K}\mathcal{O}^{\mathrm{raw-}K}_i} = 1
\end{equation}
and 
\begin{equation}
    \mathcal{O}^J_i = \frac{\mathcal{O}^{\mathrm{raw-}J}_i}{\sum_{K}\mathcal{O}^{\mathrm{raw-}K}_i} \ge 0.
\end{equation}

As stated in the main text, the proposed normalization operation is robust even in the presence of tiny components. Here, we elaborate on the underlying rationale through a systematic examination of four potential confounding factors associated with such components:
\begin{enumerate}
    \item Null Products: Although individual variable products may frequently evaluate to zero, the definition of the correlation function as a statistical mean naturally mitigates this issue.
    \item Signal-to-noise Ratio: The intrinsic noise itself—arising from Markov jump fluctuations mapped to channels—does not degrade the Signal-to-noise ratio; what influences the signal-to-noise ratio is the statistics of rare events. Here, it does not affect the validity of the results, only the efficiency in applications. (Note: Instrumental additive and counting noises are omitted from this theoretical scope but must be suppressed in empirical tests.)
    \item Covariance: The normalization process in tiny-component limits does not strip away critical covariance data when the denominator (normalization constant) remains unchanged, since it just scales by the square of the denominator when normalizing.
    \item More Permutations: Tiny components introduce a wider repertoire of channel trajectories (tiny components mean more components), yet this does not result in inflated likelihoods of sporadic events. Through extended and repeated measurements, frequency-based probability estimation remains valid.
\end{enumerate}
However, although it can still be applied within this framework, it is not recommended for thermodynamic inference. This is because, in practice, it is often uncertain whether the proliferation of channels might obscure the underlying microscopic dynamics. Moreover, when a quantity approaches zero, its relative change is typically difficult to measure.

Even when the raw sum varies from state to state, normalization can still be applied ``forcibly'': $\mathcal{O}'^{J}_i(t) = \mathcal{O}^{\mathrm{raw-}J}_i/\sum_{K}\mathcal{O}^{\mathrm{raw-}K}_i$, though this results in a time-dependent quantity. In the rest of the Supplemental Material, it can be found that our conclusion remains formally unchanged even when $\mathcal{O}'^{J}_i(t)$ is time-dependent. However, requiring the sum to remain nonzero constitutes a stringent condition outside the composition-like framework. One may also manually consider time points of zero-sum as non-contributory, preserving its broad practicality. To preclude such exceptions from confounding the theoretical analysis, we just merely note this feasibility here and do not present it as a main result.

Note that ``observables'' refer not only to empirical data from physical experiments but also to results from simulations. Provided the underlying microscopic dynamics are Markovian under specific approximations, state variables derived from molecular dynamics simulations are considered valid observables within our framework. 
For example, the normalized observables can be the fractions of the molecular motor's different conformation sets in a simulation, provided the motor dynamics are Markovian.
However, if Markovianity appears solely in the coarse-grained Markov state models (derived from microstate coarse-graining \cite{noe2019markov}), our framework will then remain mathematically self-consistent only when local detailed balance of the this scale's dynamics is additionally satisfied \cite{falasco2025macroscopic, schwarz2024mind}.

\section{Reconstruction: Lower Bounds of Entropy Production Rate}
From the perspective of information theory, $\sigma_{\left [ 0,t \right ]}$ can be expressed as 
    \begin{equation}
        \sigma_{\left [ 0,t \right ]} = D_{\mathrm{KL}}(\mathbb{P}_\gamma\|\mathbb{P}^\dagger_\gamma),
    \end{equation}
where the definition of Kullback-Leibler divergence is $D_{\mathrm{KL}}(P\|Q)=\sum_{i}p_i\log\frac{p_i}{q_i}$ with $P=\{p_i\}, Q=\{q_i\}$ \cite{polyanskiy2025information}, and $\mathbb{P}_\gamma$ denotes the probability measure of the Markov process $\gamma$ on the trajectory space
\begin{equation}
    \begin{split}
        \Omega_{\mathrm{tot}} = \bigcup_{K=0}^{\infty}&\{ (i_0,i_1,i_2,\dots,i_K;\tau_1,\tau_2,\dots,\tau_K) | i_k \in \left \{1,\dots,N_i \right \} , \tau_k>0, \sum_{k=1}^K\tau_k \le \Delta t \}.
    \end{split}
\end{equation}

And the reconstruction (i.e., its estimation counterpart) $\sigma^{\mathrm{est-}n}_{\Delta t, \{q_k\}_{k=0}^n,\left [ 0,t \right ]}$ can be rewritten as
\begin{equation}
    \begin{split}
        \sigma^{\mathrm{est-}n}_{\Delta t, \{q_k\}_{k=0}^n,\left [ 0, t \right ]}=\int_0^{t} \mathrm{d}\tau\ \dot\sigma^{\mathrm{est-}n}_{\Delta t, \{q_k\}_{k=0}^n}(\tau) = \frac{1}{\Delta t} \int_0^{t}\sum_{\{ J_k \}_{k=0}^n}\mathrm{d}\tau\ C^{J_n,\dots,J_1,J_0}_{\tau,\Delta t,\{q_k\}_{k=0}^n}\ln\frac{C^{J_n,\dots,J_1,J_0}_{\tau,\Delta t,\{q_k\}_{k=0}^n}}{C^{J_0,J_1,\dots,J_n}_{\tau,\Delta t,\{1-q_{n-k}\}_{k=0}^n}}
    \end{split}
\end{equation}
Without loss of generality, $C^{J_n,\dots,J_1,J_0}_{t,\Delta t,\{q_k\}_{k=0}^n}$ is invariant under time translations in non-equilibrium steady states (NESS); consequently, the expression simplifies to
\begin{equation}
    \begin{split}
        &\sigma^{\mathrm{est-}n}_{\Delta t, \{q_k\}_{k=0}^n,\left [ 0,t \right ]} = \frac{t}{\Delta t} \sum_{\{ J_k \}_{k=0}^n} C^{J_n,\dots,J_1,J_0}_{0,\Delta t,\{q_k\}_{k=0}^n}\ln\frac{C^{J_n,\dots,J_1,J_0}_{0,\Delta t,\{q_k\}_{k=0}^n}}{C^{J_0,J_1,\dots,J_n}_{0,\Delta t,\{1-q_{n-k}\}_{k=0}^n}} \\
        =&\frac{t}{\Delta t} \sum_{\{ J_k \}_{k=0}^n} P (\omega^{J_n,\dots,J_1,J_0}_{0,\Delta t,\{q_k\}_{k=0}^n})\ln\frac{P (\omega^{J_n,\dots,J_1,J_0}_{0,\Delta t,\{q_k\}_{k=0}^n})}{P^\dagger (\omega^{J_n,\dots,J_1,J_0}_{0,\Delta t,\{q_k\}_{k=0}^n})}= \frac{t}{\Delta t} D_{\mathrm{KL}}(P_{\omega^{J_n,\dots,J_1,J_0}_{0,\Delta t,\{q_k\}_{k=0}^n}}\|P^\dagger_{\omega^{J_n,\dots,J_1,J_0}_{0,\Delta t,\{q_k\}_{k=0}^n}}),
        \label{KL}
    \end{split}
\end{equation}
where $P_{\omega^{J_n,\dots,J_1,J_0}_{0,\Delta t,\{q_k\}_{k=0}^n}}$ denotes the probability of the event $\omega^{J_n,\dots,J_1,J_0}_{t,\Delta t,\{q_k\}_{k=0}^n}$ in the space $\Omega_{n+1-\mathrm{order}} = \left \{ 0',1',\dots,N_J' \right \} ^{n+1}$.

Since each event $\omega^{J_n,\dots,J_1,J_0}_{0,\Delta t,\{q_k\}_{k=0}^n} \in \Omega_{n+1-\mathrm{order}}$ is the image of a trajectory $\gamma \in \Omega_{\mathrm{tot}}$ under the random mapping, the data processing inequality (DPI) for the Kullback–Leibler (KL) divergence \cite{polyanskiy2025information} yields
\begin{equation}
    \sigma_{\left [ 0, t \right ]} = \frac{t}{\Delta t}D_{\mathrm{KL}}(\mathbb{P}_\gamma\|\mathbb{P}^\dagger_\gamma) \ge \frac{t}{\Delta t} D_{\mathrm{KL}}(P_{\omega^{J_n,\dots,J_1,J_0}_{0,\Delta t,\{q_k\}_{k=0}^n}}\|P^\dagger_{\omega^{J_n,\dots,J_1,J_0}_{0,\Delta t,\{q_k\}_{k=0}^n}}) = \sigma^{\mathrm{est-}n}_{\Delta t, \{q_k\}_{k=0}^n,\left [ 0, t \right ]}.
\end{equation}

Assuming stationary and finite Kullback–Leibler divergences, dividing by $\tau>0$ yields the rate bound 
\begin{equation}
    \dot\sigma(t) \ge \dot\sigma^{\mathrm{est-}n}_{\Delta t, \{q_k\}_{k=0}^n}(t)
    \label{14}
\end{equation}
holding for all $t$, thereby establishing the reconstruction as a lower bound on the entropy production rate (EPR) stated in the main text.

\section{Hierarchy of Tighter Bounds}
Assuming that readouts are indexed by slice and independent conditional on the microscopic trajectory, even at coincident times. Adopting a probability-theoretic viewpoint, $\dot\sigma^{\mathrm{est-}n+1}_{\Delta t, \{q_k\}_{k=0}^{n+1}}(t)$ can be recast as follows:
\begin{equation}
\begin{split}
    &\dot\sigma^{\mathrm{est-}n+1}_{\Delta t, \{q_k\}_{k=0}^{n+1}}(t)=\frac{1}{\Delta t} \sum_{\{ J_k \}_{k=0}^{n+1}} C^{J_{n+1},J_n,\dots,J_1,J_0}_{t,\Delta t,\{q_k\}_{k=0}^{n+1}}\ln\frac{C^{J_{n+1},J_n,\dots,J_1,J_0}_{t,\Delta t,\{q_k\}_{k=0}^{n+1}}}{C^{J_0,J_1,\dots,J_n,J_{n+1}}_{t,\Delta t,\{q_{n+1-k}\}_{k=0}^{n+1}}}\\
    =&\frac{1}{\Delta t}\sum_{\{ J_k \}_{k=0}^{n+1}} P (\omega^{J_{n+1},J_n,\dots,J_1,J_0}_{t,\Delta t,\{q_k\}_{k=0}^{n+1}})\ln\frac{P (\omega^{J_{n+1},J_n,\dots,J_1,J_0}_{t,\Delta t,\{q_k\}_{k=0}^{n+1}})}{P^\dagger (\omega^{J_{n+1},J_n,\dots,J_1,J_0}_{t,\Delta t,\{q_k\}_{k=0}^{n+1}})}\\
    =&\frac{1}{\Delta t}\sum_{\{ J_k \}_{k=0}^{n+1}}P (\omega^{J_{n+1},J_n,\dots,J_1,J_0}_{t,\Delta t,\{q_k\}_{k=0}^{n+1}})\ln\frac{P (\omega^{J_j}_{t,\Delta t,q_j} |\omega^{J_{n+1},\dots,J_{j+1},J_{j-1},\dots,J_0}_{t,\Delta t,\{q_k\}_{k=0}^{j-1}\cup\{q_k\}_{k=j+1}^{n+1}}) P (\omega^{J_{n+1},\dots,J_{j+1},J_{j-1},\dots,J_0}_{t,\Delta t,\{q_k\}_{k=0}^{j-1}\cup\{q_k\}_{k=j+1}^{n+1}})}{P^\dagger (\omega^{J_j}_{t,\Delta t,q_j}|\omega^{J_{n+1},\dots,J_{j+1},J_{j-1},\dots,J_0}_{t,\Delta t,\{q_k\}_{k=0}^{j-1}\cup\{q_k\}_{k=j+1}^{n+1}}) P^\dagger (\omega^{J_{n+1},\dots,J_{j+1},J_{j-1},\dots,J_0}_{t,\Delta t,\{q_k\}_{k=0}^{j-1}\cup\{q_k\}_{k=j+1}^{n+1}})}\\
    =&\frac{1}{\Delta t}[\sum_{\{ J_k \}_{k=0}^{n+1}}P (\omega^{J_{n+1},J_n,\dots,J_1,J_0}_{t,\Delta t,\{q_k\}_{k=0}^{n+1}})\ln\frac{P (\omega^{J_{n+1},\dots,J_{j+1},J_{j-1},\dots,J_0}_{t,\Delta t,\{q_k\}_{k=0}^{j-1}\cup\{q_k\}_{k=j+1}^{n+1}})}{P^\dagger (\omega^{J_{n+1},\dots,J_{j+1},J_{j-1},\dots,J_0}_{t,\Delta t,\{q_k\}_{k=0}^{j-1}\cup\{q_k\}_{k=j+1}^{n+1}})}\\
    &~~~~~+\sum_{\{ J_k \}_{k=0}^{n+1}}P (\omega^{J_{n+1},J_n,\dots,J_1,J_0}_{t,\Delta t,\{q_k\}_{k=0}^{n+1}})\ln\frac{P (\omega^{J_j}_{t,\Delta t,q_j}|\omega^{J_{n+1},\dots,J_{j+1},J_{j-1},\dots,J_0}_{t,\Delta t,\{q_k\}_{k=0}^{j-1}\cup\{q_k\}_{k=j+1}^{n+1}})}{P^\dagger (\omega^{J_j}_{t,\Delta t,q_j}|\omega^{J_{n+1},\dots,J_{j+1},J_{j-1},\dots,J_0}_{t,\Delta t,\{q_k\}_{k=0}^{j-1}\cup\{q_k\}_{k=j+1}^{n+1}}) }]\\
    =&\frac{1}{\Delta t}\{\sum_{\{ J_k \}_{k=0}^{j-1}\cup\{J_k\}_{k=j+1}^{n+1}}P (\omega^{J_{n+1},\dots,J_{j+1},J_{j-1},\dots,J_0}_{t,\Delta t,\{q_k\}_{k=0}^{j-1}\cup\{q_k\}_{k=j+1}^{n+1}})\ln\frac{P (\omega^{J_{n+1},\dots,J_{j+1},J_{j-1},\dots,J_0}_{t,\Delta t,\{q_k\}_{k=0}^{j-1}\cup\{q_k\}_{k=j+1}^{n+1}})}{P^\dagger (\omega^{J_{n+1},\dots,J_{j+1},J_{j-1},\dots,J_0}_{t,\Delta t,\{q_k\}_{k=0}^{j-1}\cup\{q_k\}_{k=j+1}^{n+1}})}\\
    &~~~~~+\sum_{\{ J_k \}_{k=0}^{j-1}\cup\{J_k\}_{k=j+1}^{n+1}} [P (\omega^{J_{n+1},\dots,J_{j+1},J_{j-1},\dots,J_0}_{t,\Delta t,\{q_k\}_{k=0}^{j-1}\cup\{q_k\}_{k=j+1}^{n+1}})\\
    &~~~~~~~~~~~~~~~~~~~~~~~~~~~~\times\sum_{J_j} P (\omega^{J_j}_{t,\Delta t,q_j}|\omega^{J_{n+1},\dots,J_{j+1},J_{j-1},\dots,J_0}_{t,\Delta t,\{q_k\}_{k=0}^{j-1}\cup\{q_k\}_{k=j+1}^{n+1}})\ln\frac{P (\omega^{J_j}_{t,\Delta t,q_j}|\omega^{J_{n+1},\dots,J_{j+1},J_{j-1},\dots,J_0}_{t,\Delta t,\{q_k\}_{k=0}^{j-1}\cup\{q_k\}_{k=j+1}^{n+1}})}{P^\dagger (\omega^{J_j}_{t,\Delta t,q_j}|\omega^{J_{n+1},\dots,J_{j+1},J_{j-1},\dots,J_0}_{t,\Delta t,\{q_k\}_{k=0}^{j-1}\cup\{q_k\}_{k=j+1}^{n+1}}) }]\}\\
    =& \dot\sigma^{\mathrm{est-}n}_{\Delta t, \{q_k\}_{k=0}^{j-1}\cup\{q_k\}_{k=j+1}^{n+1}}(t)+\frac{1}{\Delta t}[\sum_{\{ J_k \}_{k=0}^{j-1}\cup\{J_k\}_{k=j+1}^{n+1}} P (\omega^{J_{n+1},\dots,J_{j+1},J_{j-1},\dots,J_0}_{t,\Delta t,\{q_k\}_{k=0}^{j-1}\cup\{q_k\}_{k=j+1}^{n+1}})\\
    &~~~~~~~~~~~~~~~~~~~~~~~~~~~~~~\times\sum_{J_j} P (\omega^{J_j}_{t,\Delta t,q_j}|\omega^{J_{n+1},\dots,J_{j+1},J_{j-1},\dots,J_0}_{t,\Delta t,\{q_k\}_{k=0}^{j-1}\cup\{q_k\}_{k=j+1}^{n+1}})\ln\frac{P (\omega^{J_j}_{t,\Delta t,q_j}|\omega^{J_{n+1},\dots,J_{j+1},J_{j-1},\dots,J_0}_{t,\Delta t,\{q_k\}_{k=0}^{j-1}\cup\{q_k\}_{k=j+1}^{n+1}})}{P^\dagger (\omega^{J_j}_{t,\Delta t,q_j}|\omega^{J_{n+1},\dots,J_{j+1},J_{j-1},\dots,J_0}_{t,\Delta t,\{q_k\}_{k=0}^{j-1}\cup\{q_k\}_{k=j+1}^{n+1}}) }].\\
\end{split}
\end{equation}

The final expression contains a non-negative conditional Kullback–Leibler divergence. Hence, for $1\le j\le n$, we obtain the inequality:
\begin{equation}
    \dot\sigma^{\mathrm{est-}n+1}_{\Delta t, \{q_k\}_{k=0}^{n+1}}(t) \ge \dot\sigma^{\mathrm{est-}n}_{\Delta t, \{q_k\}_{k=0}^{j-1}\cup\{q_k\}_{k=j+1}^{n+1}}.
    \label{plug-in}
\end{equation}

Analogous to yet distinct from Refs.  \cite{roldan2010estimating, roldan2012entropy, roldan2021quantifying, kapustin2024utilizing}, the probability distribution here pertains to events mapped and extracted from the Markovian trajectories, i.e. an isomorphic construction of correlation functions, rather than to the measured trajectories themselves.

\section{How Far from $\dot\sigma^{\mathrm{est-}n}_{\Delta t, \{q_k\}_{k=0}^n}(t)$ to True Entropy Production Rate.}

\subsection{Definitions and Non-negativity of $\dot \sigma^\mathrm{oc}_{\Delta t}(t)$, $\dot \sigma^\mathrm{hid}_{\Delta t}(t)$ and $\dot \sigma^\mathrm{amb}_{\Delta t}(t)$}

Consider a stationary finite-state Markov jump process with a finite entropy production rate. Let
\begin{equation}
    \Pi_m=\{t=t_0^{(m)}<\cdots<t_{n_m}^{(m)}=t+\Delta t\}
\end{equation}
be nested temporal partitions satisfying
\begin{equation}
    \Pi_m\subseteq\Pi_{m+1},\qquad
    \|\Pi_m\|=\max_k(t_k^{(m)}-t_{k-1}^{(m)})\to0.
\end{equation}
Their union
\begin{equation}
    D=\bigcup_{m=1}^{\infty}\Pi_m
\end{equation}
is a countable dense set of observation times. Conditional on the microscopic trajectory $\gamma$, the readouts $\{J_s:s\in D\}$ are independent, with probabilities $p_{J\gets i(s)}^{\mathrm{map}}=\mathcal O_{i(s)}^J$. The same observation kernel is used under the forward and stationary reversed dynamics. We denote the joint laws of the microscopic trajectory and auxiliary readouts by $\mathbb P$ and $\mathbb P^\dagger$.

For a finite set $S=\{s_0<\cdots<s_n\}$, write
\begin{equation}
    J_S=(J_{s_0},\ldots,J_{s_n}).
\end{equation}
We define the dense-record observation-channel KL divergence by
\begin{equation}
    \sigma^{\mathrm{oc}}_{t,\Delta t}
    \coloneqq D_{\mathrm{KL}}\left(
        \mathbb P_{J_D}\middle\|\mathbb P^\dagger_{J_D}\right),
    \label{dense-record-kl}
\end{equation}
and the corresponding finite-window reconstruction rate by
\begin{equation}
    \dot\sigma^{\mathrm{oc}}_{\Delta t}(t)
    \coloneqq\frac{1}{\Delta t}
    D_{\mathrm{KL}}\left(
        \mathbb P_{J_D}\middle\|\mathbb P^\dagger_{J_D}\right).
    \label{def-oc-independent}
\end{equation}
Equivalently,
\begin{equation}
    \dot\sigma^{\mathrm{oc}}_{\Delta t}(t)
    =\frac{1}{\Delta t}\sup_{\Pi\Subset D}
    D_{\mathrm{KL}}\left(
        \mathbb P_{J_\Pi}\middle\|\mathbb P^\dagger_{J_\Pi}\right),
    \label{def-oc-supremum}
\end{equation}
where $\Pi\Subset D$ denotes a finite subset. Since the observed process need not be Markovian, stationarity alone does not make its finite-window KL divergence linear in $\Delta t$. Thus $\dot\sigma^{\mathrm{oc}}_{\Delta t}$ may depend on the window length.

To define an edge-resolved intermediate record, we specify an auxiliary marking rule at actual microscopic jumps. For each jump $i_{\ell-1}\to i_\ell$ at time $\tau_\ell$, draw labels $A_\ell$ and $B_\ell$ independently from $\mathcal O_{i_{\ell-1}}$ and $\mathcal O_{i_\ell}$, respectively. These draws are independent across jumps and of $J_D$, conditional on $\gamma$. Record the jump time and microscopic ordered pair when $A_\ell\ne B_\ell$:
\begin{equation*}
    E=\{(\tau_\ell,i_{\ell-1},i_\ell):A_\ell\ne B_\ell\},
    \qquad \gamma^{\mathrm{edge}}=(J_D,E).
\end{equation*}
The same marking rule is used under both dynamics. It marks only actual microscopic jumps. In deterministic lumping, it records precisely the microscopic identities of inter-channel jumps.

We then define
\begin{subequations}
\begin{align}
    \dot\sigma^{\mathrm{amb}}_{\Delta t}(t)
    &\coloneqq\frac{1}{\Delta t}\left[
    D_{\mathrm{KL}}\left(\mathbb P_{\gamma^{\mathrm{edge}}}
        \middle\|\mathbb P^\dagger_{\gamma^{\mathrm{edge}}}\right)
    -D_{\mathrm{KL}}\left(\mathbb P_{J_D}
        \middle\|\mathbb P^\dagger_{J_D}\right)\right],\\
    \dot\sigma^{\mathrm{hid}}_{\Delta t}(t)
    &\coloneqq\frac{1}{\Delta t}\left[
    D_{\mathrm{KL}}\left(\mathbb P_\gamma
        \middle\|\mathbb P^\dagger_\gamma\right)
    -D_{\mathrm{KL}}\left(\mathbb P_{\gamma^{\mathrm{edge}}}
        \middle\|\mathbb P^\dagger_{\gamma^{\mathrm{edge}}}\right)\right].
    \label{kl-minus}
\end{align}
\end{subequations}
Here $\dot\sigma^{\mathrm{hid}}_{\Delta t}$ is the irreversibility still hidden after edge marking, while $\dot\sigma^{\mathrm{amb}}_{\Delta t}$ is the additional irreversibility lost when the marks are discarded. For stochastic mappings, this split depends on the stated marking convention; the sum of the two terms does not.

The stochastic mapping $\gamma\mapsto\gamma^{\mathrm{edge}}$ and the projection $\gamma^{\mathrm{edge}}\mapsto J_D$ give, by the data-processing inequality,
\begin{equation*}
    D_{\mathrm{KL}}(\mathbb P_\gamma\|\mathbb P^\dagger_\gamma)
    \ge D_{\mathrm{KL}}(\mathbb P_{\gamma^{\mathrm{edge}}}
        \|\mathbb P^\dagger_{\gamma^{\mathrm{edge}}})
    \ge D_{\mathrm{KL}}(\mathbb P_{J_D}\|\mathbb P^\dagger_{J_D}).
\end{equation*}
Consequently,
\begin{equation}
    \dot\sigma^{\mathrm{amb}}_{\Delta t}(t)\ge0,
    \qquad \dot\sigma^{\mathrm{hid}}_{\Delta t}(t)\ge0.
    \label{amb/hid>0}
\end{equation}
Using $D_{\mathrm{KL}}(\mathbb P_\gamma\|\mathbb P^\dagger_\gamma)=\Delta t\,\dot\sigma$ in NESS, we obtain
\begin{equation*}
    \dot\sigma
    =\dot\sigma^{\mathrm{oc}}_{\Delta t}
    +\dot\sigma^{\mathrm{hid}}_{\Delta t}
    +\dot\sigma^{\mathrm{amb}}_{\Delta t}.
\end{equation*}

\subsection{Why $\dot{\sigma}^{\mathrm{est}}(t)$ Converges to $\dot{\sigma}^{\mathrm{oc}}(t)$ when $\sup r_k\to0$}

We first prove the result for the nested partitions of Sec. S4.1 and then extend it to arbitrary partitions whose mesh tends to zero.

By stationarity, we set $t=0$ without loss of generality. 
For the partition
\begin{equation}
    \Pi_m
    =
    \left\{
        t+q_0^{(m)}\Delta t,\ldots,
        t+q_{n_m}^{(m)}\Delta t
    \right\},
\end{equation}
define
\begin{equation}
    r_k^{(m)}=q_k^{(m)}-q_{k-1}^{(m)}.
\end{equation}
The physical mesh size is
\begin{equation}
    \|\Pi_m\|
    =
    \Delta t \cdot (\max_k r_k^{(m)}).
\end{equation}
For fixed $\Delta t$, the conditions
$\|\Pi_m\|\to0$ and $\sup_k r_k^{(m)}\to0$ are equivalent.

Let information sets be
\begin{equation}
    \mathcal{F}_m
    =
    \sigma\{J_s:s\in\Pi_m\},
    \qquad
    \mathcal{F}_D
    =
    \sigma\{J_s:s\in D\}.
\end{equation}
Because the partitions are nested,
\begin{equation}
    \mathcal{F}_m\subseteq\mathcal{F}_{m+1},
    \qquad
    \sigma\left(\bigcup_m\mathcal{F}_m\right)
    =
    \mathcal{F}_D.
\end{equation}
The observations at previously included times are retained when the partition is refined, while observations at newly added times are drawn independently conditional on the microscopic trajectory.

The reconstruction associated with $\Pi_m$ is
\begin{equation}
    \dot{\sigma}^{\mathrm{est}}_{\Pi_m}(t)
    =
    \frac{1}{\Delta t}
    D_{\mathrm{KL}}
    \left(
        \mathbb{P}|_{\mathcal{F}_m}
        \middle\|
        \mathbb{P}^{\dagger}|_{\mathcal{F}_m}
    \right).
    \label{partition-estimator-kl}
\end{equation}
Here $\mathbb P$ and $\mathbb P^\dagger$ are joint laws of the microscopic trajectory and readouts on a common space. Conditional on the trajectory, the readouts are independent and use the same state-to-observable mapping. The data-processing inequality immediately gives
\begin{equation}
    \dot{\sigma}^{\mathrm{est}}_{\Pi_m}(t)
    \le
    \dot{\sigma}^{\mathrm{est}}_{\Pi_{m+1}}(t).
    \label{partition-monotonicity}
\end{equation}

We next identify the limit. Assume
$\mathbb{P}\ll\mathbb{P}^{\dagger}$ and that the relevant KL divergence is
finite. Let
\begin{equation}
    L=\frac{\mathrm{d}\mathbb{P}}
            {\mathrm{d}\mathbb{P}^{\dagger}},
    \qquad
    L_m=
    \mathbb{E}_{\mathbb{P}^{\dagger}}
    \left[L\mid\mathcal{F}_m\right].
\end{equation}
Then $\{L_m\}$ is a non-negative
$\mathbb{P}^{\dagger}$-martingale and
\begin{equation}
    D_{\mathrm{KL}}
    \left(
        \mathbb{P}|_{\mathcal{F}_m}
        \middle\|
        \mathbb{P}^{\dagger}|_{\mathcal{F}_m}
    \right)
    =
    \mathbb{E}_{\mathbb{P}^{\dagger}}
    \left[L_m\ln L_m\right].
\end{equation}
By martingale convergence and the convergence theorem for relative entropy
on increasing sigma-algebras,
\begin{equation}
    \lim_{m\to\infty}
    D_{\mathrm{KL}}
    \left(
        \mathbb{P}|_{\mathcal{F}_m}
        \middle\|
        \mathbb{P}^{\dagger}|_{\mathcal{F}_m}
    \right)
    =
    D_{\mathrm{KL}}
    \left(
        \mathbb{P}|_{\mathcal{F}_D}
        \middle\|
        \mathbb{P}^{\dagger}|_{\mathcal{F}_D}
    \right).
    \label{relative-entropy-filtration-limit}
\end{equation}
Combining Eqs.~\eqref{partition-estimator-kl} and
\eqref{relative-entropy-filtration-limit}, we obtain
\begin{equation}
    \lim_{\substack{m\to\infty\\\sup_k r_k^{(m)}\to0}}
    \dot{\sigma}^{\mathrm{est}}_{\Pi_m}(t)
    =
    \dot{\sigma}^{\mathrm{oc}}_{\Delta t}(t).
    \label{estimator-to-dense-oc}
\end{equation}

Now let $\widehat\Pi_m$ be any sequence of partitions with $\|\widehat\Pi_m\|\to0$, not necessarily nested. 
For a finite set of observation times $F$, write $d(F) \coloneqq D_{\rm KL}(\mathbb P_{J_F}\|\mathbb P^\dagger_{J_F})/\Delta t$.
Under both dynamics, the finite-dimensional readout laws depend continuously on the ordered observation times as long as these times remain distinct, by continuity of the microscopic transition probabilities.
For any finite $F \subseteq D$, choose distinct points of $\widehat\Pi_m$ converging to the corresponding points of $F$. 
Lower semicontinuity of relative entropy and the data-processing inequality then give $d(F)\le\liminf_m d(\widehat\Pi_m)$. 
Taking the supremum over $F$ gives $\dot\sigma^{\rm oc}_{\Delta t}(t)\le\liminf_m d(\widehat\Pi_m)$ by Eq. \eqref{def-oc-supremum}. 
Conversely, for each fixed $m$, the points of $\widehat\Pi_m$ can be approximated by distinct points of $D$. 
The same lower semicontinuity and Eq. \eqref{def-oc-supremum} give $d(\widehat\Pi_m)\le\dot\sigma^{\rm oc}_{\Delta t}(t)$.
Thus Eq. \eqref{estimator-to-dense-oc} also holds for non-nested partitions.

This proof does not assume that the observation labels form a piecewise-constant or càdlàg trajectory. In particular, the observed label may change between two sampling times even when the microscopic state has not changed. Specifically, for nested partitions, the convergence in Eq. \eqref{estimator-to-dense-oc} is convergence of KL divergences on an increasing family of observation sigma-algebras, rather than pathwise convergence of an interpolated channel trajectory.

\subsection{Special Cases: When it Degenerates into Prior Models}
\subsubsection{General Case of Standard Lumping Coarse-Graining}
First, consider the case in which each state maps to exactly one observable, (isomorphic to the standard framework of deterministic lumping coarse-graining in stochastic thermodynamics \cite{esposito2012stochastic, seifert2019stochastic}):
\begin{equation}
    p_{J \gets i}^\mathrm{map}  = \mathcal{O}^J_i(t) = \mathbb{I}_{A_J}(i),
\end{equation}
where $A_J = \{\, i \mid i \mapsto J \,\}$, the sets $A_J$ are mutually disjoint, and $\mathbb{I}_{A_J}(i)$ is the indicator function.
By Eq. \eqref{amb/hid>0}, the ambiguity contribution is nonnegative:
\begin{equation}
    \dot{\sigma}^{\rm amb}_{\Delta t}(t) \ge 0 .
\end{equation}
When several microscopic transitions produce the same observed channel change, the edge marks can carry information about the direction of time that is absent from $J_D$.

\subsubsection{Restricted Transition Case of Standard Lumping Coarse-Graining}
If each pair of distinct observation channels is connected by at most one microscopic edge, the dense channel record $J_D$ determines the times of all inter-channel jumps and the microscopic edge involved in each jump, almost surely. Thus $E$ is determined by $J_D$, and Eq. \eqref{kl-minus} gives
\begin{equation}
    \dot{\sigma}^{\rm amb}_{\Delta t}(t) = 0 .
\end{equation}

\subsection{Limit Result: Microstate Distinguishability Requires Only Observable Differences}
\begin{figure}[!bth]
    \centering
    \includegraphics[width=1\columnwidth, keepaspectratio]{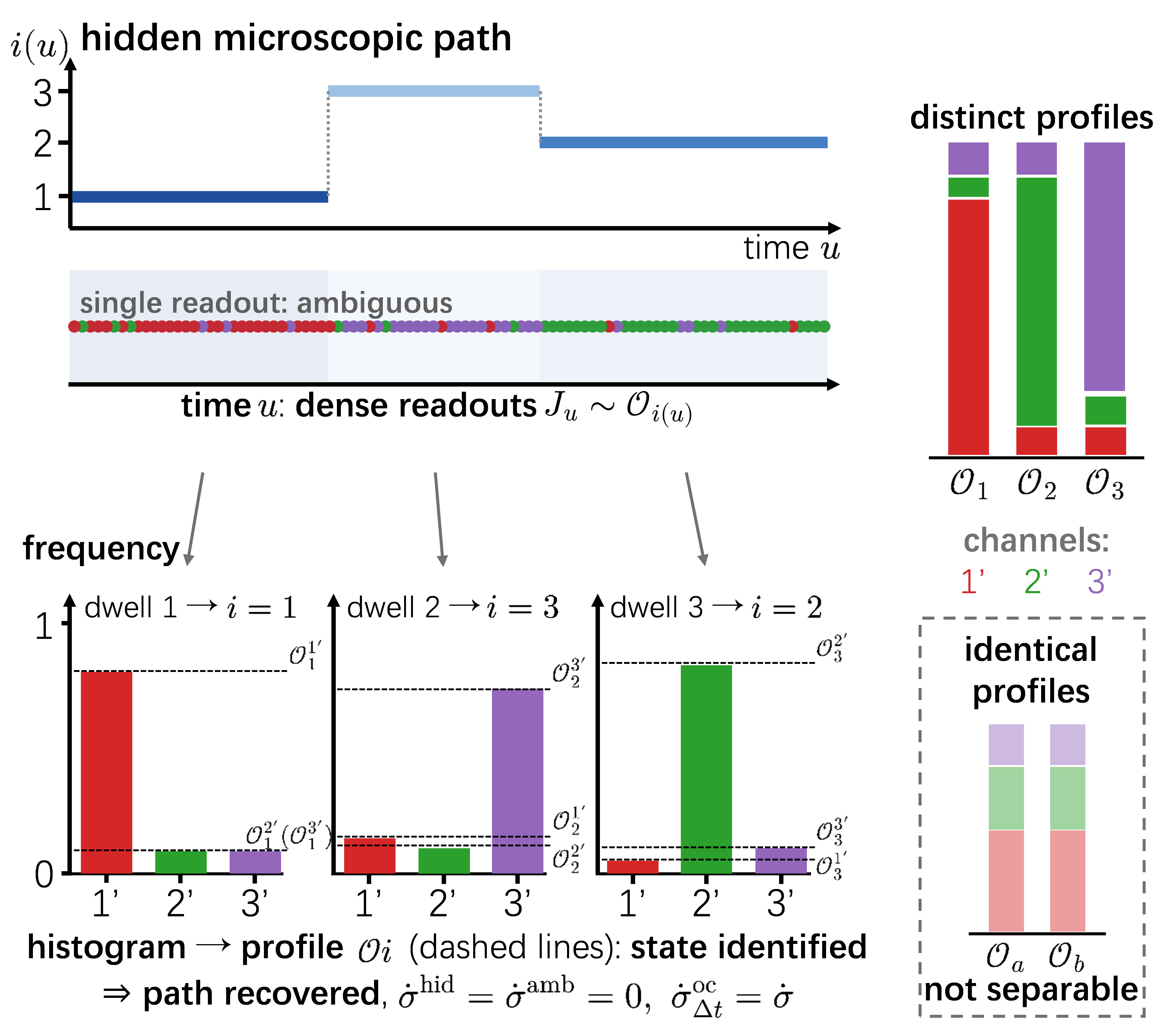}
    \caption{Mechanism underlying Eq. (12) of the main text (schematic). Top: a microscopic path visiting states 1, 3, and 2 in sequence. Middle: dense readout; each dot is a channel label extracted from the profile of the current state, so a single point is ambiguous. Bottom: during each dwell, the channel histogram converges to the profile of that state and identifies it; the dense record therefore recovers the path, giving $\dot\sigma^{\mathrm{oc}}_{\Delta t}=\dot\sigma$. States with identical characteristic profiles (inset) cannot be distinguished.}
    \label{Chernoff}
\end{figure}

The equality $\dot\sigma^{\mathrm{oc}}_{\Delta t}=\dot\sigma$ does not require a one-to-one state--channel correspondence. It holds for a stationary finite-state, non-explosive Markov jump process with a finite EPR $\dot\sigma$ when the observation profiles are pairwise distinct (illustrated schematically in the Fig. \ref{Chernoff}):
\begin{equation}
    \forall i_1\ne i_2,\qquad \exists J:
    \mathcal O_{i_1}^J\ne\mathcal O_{i_2}^J,
    \label{distinct-emission-columns}
\end{equation}
where the readouts are conditionally independent as specified in Sec. S4.1.

\subsubsection{Derivation: Recovery Map}
Let $i(u)$ denote the state at time $u\in[t,t+\Delta t]$, let $\gamma\equiv i_{[t,t+\Delta t]}$ be the path in the window, and let $J_D=\{J_u\}_{u\in D}$ be the dense record, where $J_u$ is the readout at time $u$. 
Given $\gamma$, the readouts at different times are independent, and
\begin{equation}
    \Pr(J_u=J\mid\gamma)=\mathcal{O}^J_{i(u)},
\end{equation}
under both $\mathbb{P}$ and $\mathbb{P}^\dagger$ (Section S4.1). 
Since, almost surely, the path has only finitely many jumps and is right-continuous, for every $u\in[t,t+\Delta t)$ it equals $i(u)$ on some interval $[u,u+\epsilon(\gamma,u))$, whose length $\epsilon(\gamma,u)$ depends on the hidden path and is never used. 
For every rational $s\in[t,t+\Delta t)$, fix in advance and independently of the path times $d_1(s)>d_2(s)>\cdots>s$ in $D$ converging to $s$ ($D$ is dense). 
Given $\gamma$, all but finitely many of these times lie in $[s,s+\epsilon(\gamma,s))$, where the readouts are drawn from $\mathcal{O}_{i(s)}$, and the remaining readouts are indistinguishable to the observer and do not affect the averaging in the $L\to\infty$ limit. 
Therefore, the strong law of large numbers gives, almost surely,
\begin{equation}
    \hat f_J(s) \coloneqq \lim_{L\to\infty}\frac1L\sum_{\ell=1}^{L}\mathbb{I}\{J_{d_\ell(s)}=J\}=\mathcal{O}^J_{i(s)}\qquad\forall J,
    \label{eq:S-slln}
\end{equation}
and this holds simultaneously for all rational $s$, since there are countably many. 
Let $\hat\imath(s)$ be the state whose profile equals $\hat f(s)$, which is unique by condition \eqref{distinct-emission-columns} (if no such state exists, take a fixed reference state; this case has probability zero). 
Then, almost surely, $\hat\imath(s)=i(s)$ for all rational $s$, and right-continuity extends this to every $u\in[t,t+\Delta t)$ via $i(u)=\lim_{s\downarrow u,\,s\in\mathbb{Q}}\hat\imath(s)$.
The terminal state is the left limit, since the probability of a jump at the fixed time $t+\Delta t$ is zero. 
These limits define a measurable map $f$ (constructed from the limits of countably many indicator functions of readouts, with a fixed default value where it fails), such that
\begin{equation}
    i_{[t,t+\Delta t]}=f(J_D)\quad\text{almost surely}.
    \label{eq:S-recovery}
\end{equation}
This map uses no hidden information; its construction depends only on the kernel and the path properties above. 
These properties also hold for the stationary reverse process, also a Markov jump process with generator $K^\dagger$ [Eq. \eqref{reverse-generator}] and is observed through the same kernel. 
Therefore, under $\mathbb{P}$ and $\mathbb{P}^\dagger$, Eq. \eqref{eq:S-recovery} holds with the same $f$.

Applying the data-processing inequality to the recovery map gives
\begin{equation*}
    D_{\mathrm{KL}}\left(\mathbb P_{J_D}
        \middle\|\mathbb P^\dagger_{J_D}\right)
    \ge D_{\mathrm{KL}}\left(\mathbb P_{i_{[t,t+\Delta t]}}
        \middle\|\mathbb P^\dagger_{i_{[t,t+\Delta t]}}\right).
\end{equation*}
The original observation kernel gives the opposite inequality, because it is the same under both dynamics. Thus
\begin{equation}
    D_{\mathrm{KL}}\left(\mathbb P_{J_D}
        \middle\|\mathbb P^\dagger_{J_D}\right)
    =D_{\mathrm{KL}}\left(\mathbb P_{i_{[t,t+\Delta t]}}
        \middle\|\mathbb P^\dagger_{i_{[t,t+\Delta t]}}\right).
    \label{eq:S-kl-equal}
\end{equation}

For a stationary Markov jump process, the right-hand side equals $\Delta t\,\dot\sigma$; hence, by Eq. \eqref{def-oc-independent},
\begin{equation}
\dot\sigma^{\rm oc}_{\Delta t}=\dot\sigma
\label{eq:S-oc-equals}
\end{equation}
for every $\Delta t>0$, and we may write it as $\dot\sigma^{\mathrm{oc}}$, i.e. 
\begin{equation}
    \dot\sigma^{\mathrm{oc}} \coloneqq \dot\sigma^{\mathrm{oc}}_{\Delta t} \quad (\forall ~\Delta t).
\end{equation}
This analytical result can be verified by the convergence trend shown in Fig. \ref{h0}.

If some microstates have identical profiles, complete microscopic recovery is not guaranteed.

\subsubsection{On the Necessity of Dense Observation: More Ambiguity, More Observations for Reliable Inference}
Even if we know the recovery map and the accurate inference of EPR in the dense-observation limit, this limit is unattainable in practice. We therefore need to understand how, under different levels of ambiguity, the number of observations should be varied to achieve the desired thermodynamic inference performance. This part aims to analytically demonstrate the relationship between the ambiguity and performance.

Let $P=\{P_j\}$ and $Q=\{Q_j\}$ be two distributions on a finite channel set, respectively regarded as
two hypotheses. The Chernoff information is
\begin{equation}
C(P,Q) \coloneqq -\ln\inf_{0<\theta<1}\sum_j P_j^{\,\theta}Q_j^{\,1-\theta}
\in[0,+\infty].
\label{eq:S-chernoff-info}
\end{equation}
It is symmetric and by Hölder's inequality,
\begin{equation}
\sum_j P_j^\theta Q_j^{1-\theta}
\le\Big(\sum_jP_j\Big)^\theta\Big(\sum_jQ_j\Big)^{1-\theta}=1,
\qquad 0<\theta<1,
\label{eq:S-chernoff-positive}
\end{equation}
saturating if and only if $P=Q$. The value $+\infty$ is allowed when their supports are disjoint.

For $N$ independent labels $y_1,\dots,y_N$, all drawn either from $P$ or all from $Q$ (both known), with priors $\rho_P,\rho_Q>0$, let $P^{*,(N)}_{\rm err}$ denote the Bayes error minimized over all decision rules $\phi$:
\begin{equation}
P^{*,(N)}_{\rm err} = \min\{\rho_P\Pr(\phi=Q\mid P)+\rho_Q\Pr(\phi=P\mid Q)\}.
\end{equation}
This minimum is attained by comparing $\rho_P\prod_kP_{y_k}$ with $\rho_Q\prod_kQ_{y_k}$; when the priors are equal, this is the maximum-likelihood (ML) rule. 
For any fixed priors, the Chernoff theorem gives
\begin{equation}
\lim_{N\to\infty}-\frac1N\ln P^{*,(N)}_{\rm err}=C(P,Q)
\label{eq:S-chernoff-exponent}
\end{equation}
\cite{polyanskiy2025information}, so $C(P,Q)$ is the optimal rate for distinguishing $P$ from $Q$.
ML, which requires no priors, also attains this rate: by Markov's inequality, its error under either hypothesis is at most $\big(\sum_JP_J^{\,\theta}Q_J^{\,1-\theta}\big)^N$ for every $\theta\in(0,1)$, and hence at most $e^{-NC(P,Q)}$. 
This finite-$N$ inequality is the Chernoff bound.

For a state profile $\mathcal{O}_i$ (i.e., a column in upper right part of Fig. \ref{Chernoff}), $C(\mathcal{O}_{i_1},\mathcal{O}_{i_2})$ measures the ambiguity of the mapping between two states: the number of readings required to distinguish them reliably grows as its reciprocal. 
For the entire mapping,
\begin{equation}
C_{\min}\coloneqq\min_{i_1 \neq i_2}C(\mathcal{O}_{i_1},\mathcal{O}_{i_2})
\end{equation}
is infinite when the supports are pairwise disjoint (as in a one-to-one mapping), and zero when two states share the same profile. 
Moreover, condition \eqref{distinct-emission-columns} is equivalent to $C_{\min}>0$. 
With $N_m(I)$ number of readings $J_{u_k}$ in the dwell interval $I$ drawn independently from the same profile, ML selects the state $i$ maximizing $\prod_k\mathcal{O}_i^{J_{u_k}}$, and a union bound over the $N_i-1$ alternative states gives
\begin{equation}
P_{\rm err}(I)\le(N_i-1)\,e^{-N_m(I)\,C_{\min}}.
\label{eq:S-chernoff-bound}
\end{equation}
This bound treats the dwell interval as known, which the observer cannot do. 
It therefore does not define a recovery map and is not used in the proof, which instead uses the limiting channel histogram at each time instant.
Since $N_m(I)\to\infty$ as $\|\Pi_m\|\to0$, it shows that under dense sampling, identification within a dwell interval becomes exponentially reliable, and more slowly the smaller $C_{\min}$ is.

\section{Optimization of the Reconstruction for the Fixed $n$}
All subsequent analysis is performed under a fixed $n$.

\subsection{Prerequisites for Optimizing $\Delta t$}
\subsubsection{Case of $\Delta t \to \infty$}
First, we obtain
\begin{equation}
    \lim_{\Delta t \to \infty} \dot\sigma^{\mathrm{est-}n}_{\Delta t, \{q_k\}_{k=0}^n}(t) =0.
    \label{optimal_Delta_t-1}
\end{equation}
Assume the finite-state chain is irreducible. By mixing, for fixed, distinct $\{q_k\}$, the correlation factorizes into the product of the mean observables as $\Delta t\to\infty$:
\begin{equation}
    \begin{split}
        &\lim_{\Delta t \to \infty} \dot\sigma^{\mathrm{est-}n}_{\Delta t, \{q_k\}_{k=0}^n}(t) = \lim_{\Delta t \to \infty} \sum_{\{J_k\}_{k=0}^n} \frac{C^{J_n,\dots,J_1,J_0}_{t,\Delta t,\{q_k\}_{k=0}^n}}{\Delta t} \ln \frac{C^{J_n,\dots,J_1,J_0}_{t,\Delta t,\{q_k\}_{k=0}^n}}{C^{J_0,J_1\dots,J_n}_{t,\Delta t,\{1-q_{n-k}\}_{k=0}^n}}\\
        =& \lim_{\Delta t \to \infty} \sum_{\{J_k\}_{k=0}^n} \frac{\prod_{\{J_k\}_{k=0}^n}(p_{J_k}^{\mathrm{ss}})}{\Delta t} \ln \frac{\prod_{\{J_k\}_{k=0}^n}(p_{J_k}^{\mathrm{ss}})}{\prod_{\{J_k\}_{k=0}^n}(p_{J_{n-k}}^{\mathrm{ss}})}=0.
    \end{split}
\end{equation}
Even if some $q_k = q_{k+1}$, the same result follows by replacing the relevant factors with $S_B=\sum_i p_i^{ss}\prod_{k\in B}\mathcal O_i^{J_k}$ (where $B$ denotes the set of indices corresponding to the same $q_k$ value).

\subsubsection{Case of $\Delta t \to 0$}
Second, we consider the opposite limit $\Delta t \to 0$, which requires more careful treatment.

We set $r_k=q_k-q_{k-1}$ (where $r_k \ge 0$). To evaluate the short-time limit, we expand the transition matrix element $\exp(K r_k \Delta t)_{i_k i_{k-1}}$ up to the first order of $\Delta t$:
\begin{equation}
    \exp(K r_k \Delta t)_{i_k i_{k-1}} = \delta_{i_k, i_{k-1}} + K_{i_k i_{k-1}} r_k \Delta t + \mathrm{o}(\Delta t),
\end{equation}
where $\delta_{i_k, i_{k-1}}$ is the Kronecker delta. Substituting this into the event probability $P(\omega^{J_n,\dots,J_1,J_0}_{0,\Delta t,\{q_k\}_{k=0}^n})$ and expanding the product of the matrices, we obtain:
\begin{equation}
    \begin{split}
        P(\omega^{J_n,\dots,J_1,J_0}_{0,\Delta t,\{q_k\}_{k=0}^n}) &= \sum_{\{i_k\}_{k=0}^n} \left(\prod_{k=0}^n \mathcal{O}_{i_k}^{J_k} \right) \left[\prod_{k=1}^n \exp(K r_k \Delta t)_{i_k i_{k-1}} \right] p_{i_0}^{\mathrm{ss}} + \mathrm{o}(\Delta t) \\
        &= \sum_{\{i_k\}_{k=0}^n} \left(\prod_{k=0}^n \mathcal{O}_{i_k}^{J_k} \right) \left[\prod_{k=1}^n \left( \delta_{i_k, i_{k-1}} + K_{i_k i_{k-1}} r_k \Delta t \right) \right] p_{i_0}^{\mathrm{ss}} + \mathrm{o}(\Delta t) \\
        &= \sum_{\{i_k\}_{k=0}^n} \left(\prod_{k=0}^n \mathcal{O}_{i_k}^{J_k} \right) \left[ \prod_{k=1}^n \delta_{i_k, i_{k-1}} + \sum_{j=1}^n K_{i_j i_{j-1}} r_j \Delta t \left( \prod_{k \neq j} \delta_{i_k, i_{k-1}} \right) \right] p_{i_0}^{\mathrm{ss}} + \mathrm{o}(\Delta t).
    \end{split}
    \label{prob_expansion}
\end{equation}
In the zeroth-order term, the Kronecker deltas force all intermediate states to be identical ($i_0=i_1=\dots=i_n$). We define this zeroth-order term as the \textit{static co-observation} term $S_{\{J_k\}}$, and the first-order term as the \textit{dynamic flow} term $D_{\{J_k\}}$:
\begin{equation}
    P(\omega^{J_n,\dots,J_1,J_0}_{0,\Delta t,\{q_k\}_{k=0}^n}) = S_{\{J_k\}_{k=0}^n} + D_{\{J_k\}_{k=0}^n} \Delta t + \mathrm{o}(\Delta t),
\end{equation}
where
\begin{equation}
    \begin{split}
        S_{\{J_k\}_{k=0}^n} &\coloneqq \sum_{i_0} \left(\prod_{k=0}^n \mathcal{O}_{i_0}^{J_k} \right) p_{i_0}^{\mathrm{ss}}, \\
        D_{\{J_k\}_{k=0}^n} &\coloneqq \sum_{j=1}^n r_j \sum_{u, v} \left( \prod_{k=0}^{j-1}\mathcal{O}_{u}^{J_k} \right) \left( \prod_{k=j}^n \mathcal{O}_{v}^{J_k} \right) K_{vu} p_u^{\mathrm{ss}}.
    \end{split}
\end{equation}

Here, we should conduct a categorized discussion owing to the mathematical complexity. We state the conclusions:
\begin{itemize}
    \item If $\exists i,~ \forall \mathcal{O}_i^J \ne 0$, it follows that $S_{\{J_k\}_{k=0}^n} \ne 0$, and we will obtain $\lim_{\Delta t \to 0} \dot\sigma^{\mathrm{est-}n}_{\Delta t, \{q_k\}_{k=0}^n}(t) =0$. (Importantly, $\exists i,~ \forall \mathcal{O}_i^J \ne 0 \Rightarrow S_{\{J_k\}_{k=0}^n} \ne 0$, but $S_{\{J_k\}_{k=0}^n} \ne 0 \nRightarrow \exists i,~ \forall \mathcal{O}_i^J \ne 0$.)
    \item Otherwise, we can only obtain $\lim_{\Delta t \to 0} \dot\sigma^{\mathrm{est-}n}_{\Delta t, \{q_k\}_{k=0}^n}(t) < \infty$.
\end{itemize}
For fixed $n$ and $\{q_k\}$, analyticity of the sequence probabilities gives a power--log expansion of the KL divergence at $\Delta t=0$.
Together with $0\le D_{\mathrm{KL}}\le\dot\sigma\Delta t$ and $\dot\sigma<\infty$, this ensures a finite limit of
$D_{\mathrm{KL}}/\Delta t$.

We now derive \textbf{the former statement}.

Crucially, because scalar multiplication is commutative, the static co-observation term $S_{\{J_k\}}$ is strictly symmetric under time-reversal (i.e., reversing the observed sequence from $\{J_k\}$ to $\{J_{n-k}\}$):
\begin{equation}
    S_{\{J_{n-k}\}_{k=0}^n} = \sum_{i_0} \left(\prod_{k=0}^n \mathcal{O}_{i_0}^{J_{n-k}}\right) p_{i_0}^{\mathrm{ss}} = \sum_{i_0} \left(\prod_{k=0}^n \mathcal{O}_{i_0}^{J_{k}}\right) p_{i_0}^{\mathrm{ss}} = S_{\{J_k\}_{k=0}^n},
\end{equation}
where the steady-state probability distributions of states $p_i^\mathrm{ss}$ can be obtained by solving the stationary master equation $\sum_{i'\neq i}\left[k_{i'i}p_{i}^\mathrm{ss}-k_{ii'}p_{i'}^\mathrm{ss}\right]=0$.

Under the prerequisite of $\exists i,~ \forall \mathcal{O}_i^J \ne 0$ which means that at least one state can be ``seen'' through all observation channels (observables), $S_{\{J_k\}_{k=0}^n} > 0$ holds strictly. This is typical for composition-like state observables. For example, multiple colors can all be emitted by the same state discussed in the main text. We can now substitute the expanded probability into the estimator. To simplify the notation, we denote the forward sequence as $\omega$ and the reversed sequence as $\omega^\dagger$:
\begin{equation}
    \begin{split}
        \dot\sigma^{\mathrm{est-}n}_{\Delta t, \{q_k\}_{k=0}^n}(t)\cdot\Delta t &= \sum_{\omega} P(\omega) \ln \frac{P(\omega)}{P(\omega^\dagger)} \\
        &= \sum_{\omega} \left[ S_{\omega} + D_{\omega} \Delta t + \mathrm{o}(\Delta t) \right] \ln \frac{S_{\omega} + D_{\omega} \Delta t + \mathrm{o}(\Delta t)}{S_{\omega} + D_{\omega^\dagger} \Delta t + \mathrm{o}(\Delta t)}.
    \end{split}
\end{equation}
Normalization gives
$\sum_\omega[P(\omega)-P(\omega^\dagger)]=0$.
For $S>0$, Taylor expansion yields
\begin{equation}
    (S+x)\ln\frac{S+x}{S+y}-x+y
    =\frac{(x-y)^2}{2S}+\mathrm{o}(x^2+y^2).
    \label{s0}
\end{equation}
Applying this with $S=S_\omega$,
$x=D_\omega\Delta t+\mathrm{o}(\Delta t)$ and
$y=D_{\omega^\dagger}\Delta t+\mathrm{o}(\Delta t)$ gives:
\begin{equation}
    \dot\sigma^{\mathrm{est-}n}_{\Delta t, \{q_k\}_{k=0}^n}(t)\cdot\Delta t = \left( \sum_{\omega} \frac{(D_\omega - D_{\omega^\dagger})^2}{2 S_\omega} \right) \Delta t^2 + \mathrm{o}(\Delta t^2).
\label{limit_O2}
\end{equation}
Finally, dividing by $\Delta t$ and taking the limit $\Delta t \to 0$, we rigorously prove that:
\begin{equation}
    \lim_{\Delta t \to 0} \dot\sigma^{\mathrm{est-}n}_{\Delta t, \{q_k\}_{k=0}^n}(t) = \lim_{\Delta t \to 0} \frac{O(\Delta t^2)}{\Delta t} = 0.
    \label{optimal_Delta_t-2}
\end{equation}

Physically, Eq. \eqref{limit_O2} reveals that in the strict $\Delta t \to 0$ limit, the observed dynamical irreversibility (encoded in $D_\omega - D_{\omega^\dagger}$) is completely overwhelmed by the background ambiguity of static observations (the denominator $S_\omega$). Thus, exploring finite time intervals $\Delta t_{\mathrm{opt}}$ is theoretically imperative for optimally extracting the thermodynamic bounds under ambiguous experimental mappings.

\subsubsection{The non-negativity on the whole domain of $\Delta t$}
Third, we obtain:
\begin{equation}
    \dot \sigma^{\mathrm{est-}n}_{\Delta t, \{q_k\}_{k=0}^n} (t) \ge 0.
    \label{optimal_Delta_t-3}
\end{equation}
It can be derived as follows. According to Eq. \eqref{KL},
\begin{equation}
    \begin{split}
        &\dot \sigma^{\mathrm{est-}n}_{\Delta t, \{q_k\}_{k=0}^n} (t) = \frac{\partial \sigma^{\mathrm{est-}n}_{\Delta t, \{q_k\}_{k=0}^n,[0,t]}}{\partial t} = \frac{\partial \frac{t}{\Delta t} D_{\mathrm{KL}}\left(P_{\omega^{J_n,\dots,J_1,J_0}_{0,\Delta t,\{q_k\}_{k=0}^n}}\|P^\dagger_{\omega^{J_n,\dots,J_1,J_0}_{0,\Delta t,\{q_k\}_{k=0}^n}}\right)}{\partial t} \\
        =& \frac{1}{\Delta t} D_{\mathrm{KL}}\left(P_{\omega^{J_n,\dots,J_1,J_0}_{0,\Delta t,\{q_k\}_{k=0}^n}}\|P^\dagger_{\omega^{J_n,\dots,J_1,J_0}_{0,\Delta t,\{q_k\}_{k=0}^n}}\right) \ge 0.
    \end{split}
\end{equation}

\subsection{Existence of $\Delta t_{\mathrm{opt}-n}$ and $(\{q_k\}_{k=0}^n)_{\mathrm{opt}}$}
For fixed order $n$, let $\mathcal Q_n=\{(q_0,\ldots,q_n):0=q_0\le q_1\le\cdots\le q_n=1\}$ denote the admissible domain of relative sampling times, inclusive of its boundary.
We define $F_n(\Delta t,q) = \dot\sigma^{\mathrm{est}-n}_{\Delta t,\{q_k\}_{k=0}^n}(t),~ q\in\mathcal Q_n$.

The existence of a global optimum is understood under the following regularity assumptions. First, $F_n(\Delta t,q)$ is continuous on
$(0,\infty)\times\mathcal Q_n$. Second, the short-time limit
\begin{equation}
    F_n(0,q):=\lim_{\Delta t\to 0^+}F_n(\Delta t,q)
\end{equation}
exists and gives a continuous extension of $F_n$ to $[0,\infty)\times\mathcal Q_n$. This condition includes the boundary $q_i=q_{i+1}$ and excludes singular limits depending on the path by which $(\Delta t,q)$ approaches the boundary. Third, the long-time decay is uniform with respect to the sampling time points: $\lim_{T\to\infty}\sup_{\Delta t\ge T,\;q\in\mathcal Q_n} F_n(\Delta t,q)=0 $ (i.e., uniform convergence).
This uniformity condition is needed because the relative intervals $q_k-q_{k-1}$ may themselves approach the boundary as $\Delta t$ increases.

Under these assumptions, the estimator can be continuously extended to the compactified domain $[0,\infty]\times\mathcal Q_n$ by setting
$F_n(\infty,q)=0$. Since $\mathcal Q_n$ is compact and the extended function is continuous, the Weierstrass theorem guarantees that $F_n$ attains its supremum.
Therefore, there exists at least one optimal parameter set such that
\begin{equation}
    \dot\sigma^{\mathrm{est}-n}_{\Delta t_{\mathrm{opt}-n}, (\{q_k\}_{k=0}^n)_{\mathrm{opt}}}(t) = \max_{\Delta t\ge0,\;q\in\mathcal Q_n} \dot\sigma^{\mathrm{est}-n}_{\Delta t,\{q_k\}_{k=0}^n}(t).
\end{equation}
If the maximum value is strictly positive, the optimizer cannot occur at
$\Delta t=\infty$, because $F_n(\infty,q)=0$.

(Caution: the boundary of the $\{q_k\}_{k=0}^n$ domain cannot be excluded: $q_i = q_{i+1}$ need not imply a reduction of the effective order $n$, particularly for small $\Delta t$.)

\section{Details of the Numerical Example}
\begin{figure}[!bth]
    \centering
    \includegraphics[width=0.74\columnwidth]{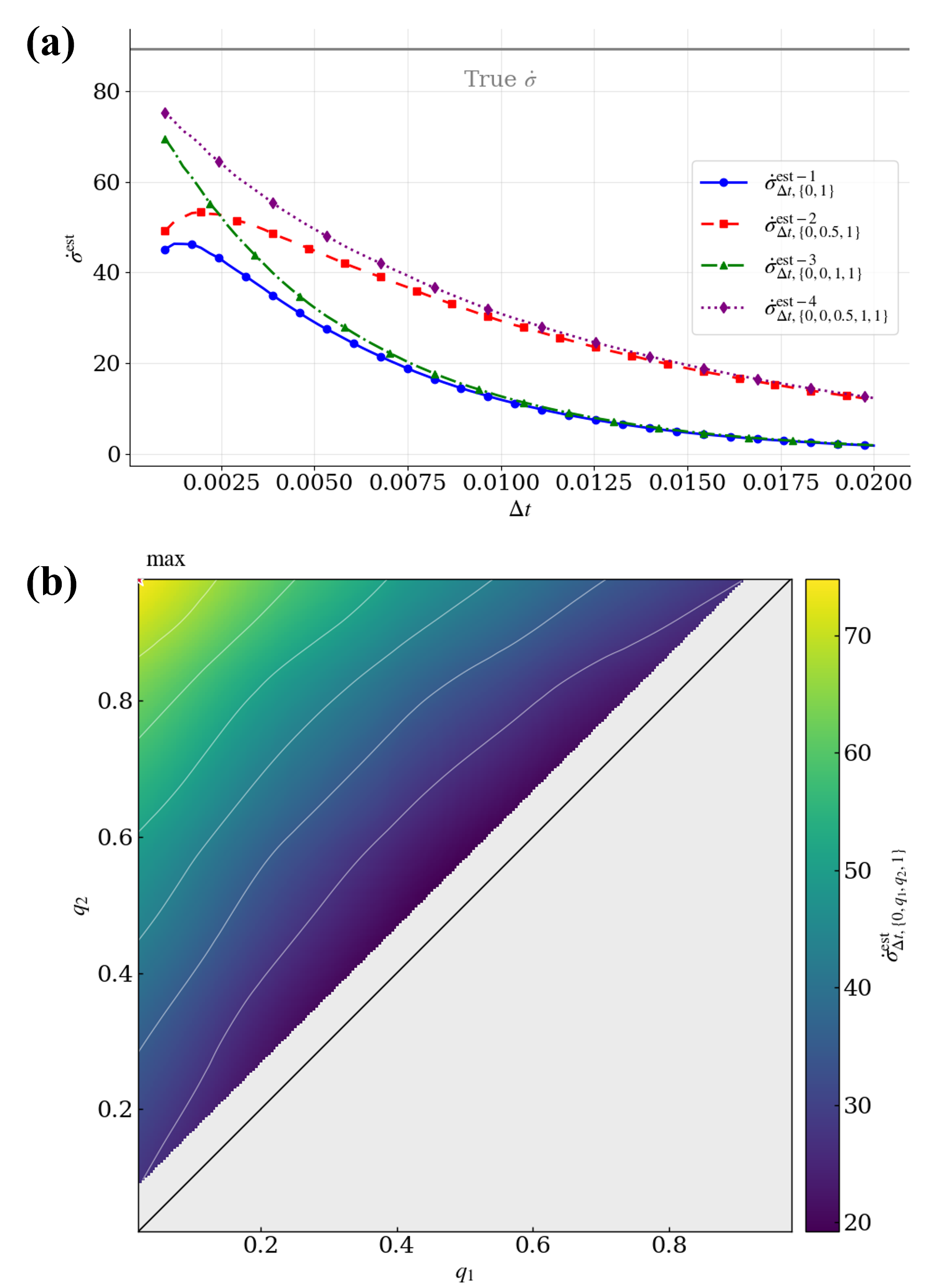}
    \caption{(a) Cross section of search for $\Delta t_{\mathrm{opt}-n}$ at $k^+/k^-=6$ and $n=1,2,3,4$. (b) Cross section of search for $(\{q_k\}_{k=0}^n)_{\mathrm{opt}}$ at $n=3$, $k^+/k^-=6$ and $\Delta t = 0.000116$. With scanning step $\Delta q_k = 0.0686$, it shows how $\dot\sigma^{\mathrm{est-}n}_{\Delta t, \{q_k\}_{k=0}^n}(t)$ changes as $q_1$ and $q_2$ vary simultaneously (the gray area represents the undesirable parameter range). Note that: all parameters in the exploration should change simultaneously, and these figures only illustrate how parameter adjustments affect the estimator.}
    \label{opt}
\end{figure}

\subsection{Continuous Model Based on Ref. \cite{song2024information}}
Following Ref.~\cite{song2024information}, distinct states may emit photons in the same color channel, while one state may emit photons in several channels. The recoloring matrix $R$ has entries $R_{Ji}=p_{J\gets i}^{\mathrm{map}}$, with $\sum_JR_{Ji}=1$. The emission profile of each state is fixed.

Assume a state-independent total photon emission rate $\mu$. Along a microscopic trajectory $i(t)$, the conditional optical intensity in color $J$ is $\mathcal I_J(t)=\mu\mathrm h\nu_JR_{J,i(t)}$, where $\mathrm h$ is Planck's constant. Its ensemble mean is
\begin{equation}
    I_J(t)=\langle\mathcal I_J(t)\rangle
    =\mu\mathrm h\nu_J\sum_iR_{Ji}p_i(t).
    \label{intensity}
\end{equation}
After converting optical power to photon-number intensity, the total conditional emission rate is $\mu$. The normalized state observables are therefore
\begin{equation}
    \mathcal O^J(t)
    =\frac{\mathcal I_J(t)}{\mu\mathrm h\nu_J}
    =R_{J,i(t)},\qquad
    \langle\mathcal O^J(t)\rangle
    =\frac{I_J(t)}{\mu\mathrm h\nu_J}=\sum_iR_{Ji}p_i(t).
\end{equation}
The multi-time correlations use these fluctuating state observables:
\begin{equation*}
    \left\langle\prod_{k=0}^{n}\mathcal O^{J_k}(t_k)\right\rangle
    =\left\langle\prod_{k=0}^{n}R_{J_k,i(t_k)}\right\rangle.
\end{equation*}
The ensemble average is taken after forming the product.

\subsection{Setup}
The three-state biomolecular process is governed by the master equation Eq. \eqref{master} with the transition matrix
\begin{equation}
    K=\begin{pmatrix} -(k^++k^-) & k^- & k^+ \\ k^+ & -(k^++k^-) & k^- \\ k^- & k^+ & -(k^++k^-) \end{pmatrix},
\end{equation}
wherein 
\begin{equation}
p_1^\mathrm{ss}=p_2^\mathrm{ss}=p_3^\mathrm{ss}=1/3.
\end{equation}
We set
\begin{equation}
    k^- = 10
\end{equation}
for all systems, so that varying $k^+$ alone changes the system dynamics.

According to stochastic thermodynamics \cite{schnakenberg1976network,seifert2012stochastic}, the cycle affinity $\mathcal{F}_\mathrm{c}$ is given by
\begin{equation}
    \mathcal{F}_\mathrm{c} = \ln\frac{(k^+)^3}{(k^-)^3} = 3 \ln\frac{k^+}{k^-},
\end{equation}
which yields the following relation to steady-state $\dot\sigma(t)$:
\begin{equation}
    \dot\sigma(t) = \sum_{\substack{i',i \\ i'\neq i}} K_{i'i}p_i^\mathrm{ss} \ln \frac{K_{i'i}p_i^\mathrm{ss}}{K_{ii'}p_{i'}^\mathrm{ss}} = (k^+-k^-)\ln\frac{k^+}{k^-} = k^- \frac{\mathcal{F}_\mathrm{c}}{3}\left[\exp\left( \frac{\mathcal{F}_\mathrm{c}}{3} \right) -1\right]
\end{equation}

\subsection{Influence of Deviation from Diagonal Matrix on Convergence Rate from Estimator to $\dot\sigma(t)$ when $n \to \infty$}
\subsubsection{Theoretical Definition and Significance of $\dot{\sigma}^{\mathrm{oc-appro}}_h$}
For a uniform sampling interval $h$, distinct from the observation-window length $\Delta t$, the forward microscopic transition matrix is
\begin{equation}
    P_h=\exp(Kh).
\end{equation}
We consider the finite irreducible model of Sec. S6.2, with stationary distribution $\pi$ and finite EPR. The stationary reversed generator is
\begin{equation}
    K^\dagger_{i'i}=K_{ii'}\frac{\pi_{i'}}{\pi_i},
    \qquad i'\ne i,
    \label{reverse-generator}
\end{equation}
with diagonal entries chosen so that each column sums to zero. The reversed sampled transition matrix is
\begin{equation}
    P_h^\dagger=\exp(K^\dagger h).
\end{equation}
At each sampling time, the observation label is independently drawn from $R_{\cdot i}$ conditional on the microscopic state. For each label $J$, define
\begin{equation}
    D_J=\operatorname{diag}(R_{J1},\ldots,R_{J N_i}).
\end{equation}
The forward probability of $\mathbf J_{0:N}=(J_0,\ldots,J_N)$ is
\begin{equation}
    \mathbb P_h(\mathbf J_{0:N})
    =\boldsymbol 1^{\mathsf T}D_{J_N}P_hD_{J_{N-1}}P_h\cdots
        D_{J_1}P_hD_{J_0}\pi,
    \label{forward-hmm-likelihood}
\end{equation}
and its probability under the reversed dynamics is
\begin{equation}
    \mathbb P_h^\dagger(\mathbf J_{0:N})
    =\boldsymbol 1^{\mathsf T}D_{J_N}P_h^\dagger D_{J_{N-1}}P_h^\dagger\cdots
        D_{J_1}P_h^\dagger D_{J_0}\pi.
    \label{reverse-hmm-likelihood}
\end{equation}

At finite $h$, define the long-time sampled observation-channel KL rate by
\begin{equation}
    \dot\sigma^{\mathrm{oc-appro}}_h
    =\lim_{N\to\infty}\frac{1}{Nh}
    D_{\mathrm{KL}}\left(\mathbb P_h(\mathbf J_{0:N})
        \middle\|\mathbb P_h^\dagger(\mathbf J_{0:N})\right).
    \label{finite-h-oc-rate}
\end{equation}
The finite irreducible hidden Markov model has stable filters at fixed $h>0$, giving the rate above. This long-time rate is distinct from the finite-window quantity in Sec. S4.1. We next establish its dense-sampling limit when the emission distributions are pairwise distinct.

For fixed $T>0$, construct a dense set $D_T\subseteq[0,T]$ as in Sec. S4.1. Let $M_h=\lfloor T/h\rfloor+1$ and
\begin{equation*}
    B_h(T)=D_{\mathrm{KL}}\left(
        \mathbb P_h(\mathbf J_{0:M_h-1})
        \middle\|\mathbb P_h^\dagger(\mathbf J_{0:M_h-1})\right).
\end{equation*}
Approximating any fixed finite subset of $D_T$ by grid points, we obtain the corresponding lower bound on $\liminf_{h\to0}B_h(T)$ by continuity of the finite-dimensional laws and lower semicontinuity of KL. Taking the supremum over these subsets and using Secs. S4.2 and S4.4 yields $\liminf_{h\to0}B_h(T)\ge T\dot\sigma$. The microscopic DPI upper bound gives the reverse inequality, so $B_h(T)\to T\dot\sigma$.

Let $\pi_{\min}=\min_i\pi_i>0$. Conditional on any preceding observation history, a reversed block likelihood is at most $1/\pi_{\min}$ times its stationary block likelihood: both are mixtures over the initial hidden state, with weights bounded above by one and stationary weights bounded below by $\pi_{\min}$. Applying the KL chain rule to consecutive blocks of $M_h$ observations therefore gives
\begin{equation*}
    \frac{B_h(T)-\ln(1/\pi_{\min})}{M_hh}
    \le\dot\sigma^{\mathrm{oc-appro}}_h\le\dot\sigma.
\end{equation*}
The forward conditioning contributes a non-negative mutual information term in this bound. Since $M_hh\to T$, taking $h\to0$ gives
\begin{equation*}
    \liminf_{h\to0}\dot\sigma^{\mathrm{oc-appro}}_h
    \ge\dot\sigma-\frac{\ln(1/\pi_{\min})}{T}.
\end{equation*}
Finally, letting $T\to\infty$ proves
\begin{equation}
    \lim_{h\to0^+}\dot\sigma^{\mathrm{oc-appro}}_h
    =\dot\sigma^{\mathrm{oc}}_{\Delta t}
    =\dot\sigma,
    \label{dense-limit-from-h}
\end{equation}
under condition~\eqref{distinct-emission-columns}. This argument connects the two limits without assuming that they commute. The superscript ``appro'' denotes the finite sampling resolution.

For $n+1$ uniformly spaced observations in a fixed window of length $\Delta t$, the sampling interval is
\begin{equation}
    h = \frac{\Delta t}{n}.
\end{equation}
Under condition \eqref{distinct-emission-columns}, both the finite-window estimator and the long-time rate $\dot\sigma_h^{\mathrm{oc\text{-}appro}}$ tend to $\dot\sigma$ as $h\to0$. 
The dependence of the long-time rate on $h$ may serve as an empirical benchmark for the finite-window estimator, but the common limit does not imply equal convergence rates.

\subsubsection{Algorithm to Get $\dot{\sigma}^{\mathrm{oc-appro}}_h$}
To evaluate Eq. \eqref{finite-h-oc-rate} without generating stochastic trajectories, we propagate a pair of forward and reversed Bayesian filters.
After observing $\mathbf{J}_{0:k}$, define
\begin{equation}
    \alpha_k(i)
    =
    \Pr(X_{kh}=i\mid\mathbf{J}_{0:k}),
    \qquad
    \beta_k(i)
    =
    \Pr^\dagger(X_{kh}=i\mid\mathbf{J}_{0:k}).
\end{equation}
Given the filter pair $(\alpha,\beta)$, the predictive probabilities of the next label are
\begin{subequations}
\begin{align}
    \ell_J(\alpha)
    &=
    \boldsymbol{1}^{\mathsf T}D_JP_h\alpha,
    \\
    \ell_J^\dagger(\beta)
    &=
    \boldsymbol{1}^{\mathsf T}D_JP_h^\dagger\beta.
\end{align}
\label{predictive-label-probabilities}
\end{subequations}
After observing $J$, the filters are updated according to
\begin{subequations}
\begin{align}
    \Phi_J(\alpha)
    &=
    \frac{D_JP_h\alpha}{\ell_J(\alpha)},
    \\
    \Phi_J^\dagger(\beta)
    &=
    \frac{D_JP_h^\dagger\beta}
         {\ell_J^\dagger(\beta)}.
\end{align}
\label{discrete-filter-updates}
\end{subequations}

Let $\mu_h(\alpha,\beta)$ denote the invariant probability measure of the filter pair generated by the forward observation process. Its transfer operator is
\begin{equation}
    (\mathcal{L}_h\mu)(A)
    =
    \int
    \sum_J
    \ell_J(\alpha)
    \mathbb{I}_A
    \left(
        \Phi_J(\alpha),
        \Phi_J^\dagger(\beta)
    \right)
    \mathrm{d}\mu(\alpha,\beta),
    \label{filter-pair-transfer-operator}
\end{equation}
where $\mathbb I_A$ is the indicator function, equal to one if the updated forward--reverse filter pair $\bigl(\Phi_J(\alpha),\Phi_J^\dagger(\beta)\bigr)$ belongs to $A$, and zero otherwise.
Once the invariant measure has been obtained, the KL rate is
\begin{equation}
    \dot{\sigma}^{\mathrm{oc-appro}}_h
    =
    \frac{1}{h}
    \int
    \sum_J
    \ell_J(\alpha)
    \ln
    \frac{\ell_J(\alpha)}
         {\ell_J^\dagger(\beta)}
    \,
    \mathrm{d}\mu_h(\alpha,\beta).
    \label{filter-pair-kl-rate}
\end{equation}
Equation~\eqref{filter-pair-kl-rate} follows from the decomposition
\begin{equation}
    \ln
    \frac{\mathbb{P}_h(\mathbf{J}_{0:N})}
         {\mathbb{P}_h^\dagger(\mathbf{J}_{0:N})}
    =
    \sum_{k=0}^{N-1}
    \ln
    \frac{\ell_{J_{k+1}}(\alpha_k)}
         {\ell_{J_{k+1}}^\dagger(\beta_k)}
    +
    O(1),
\end{equation}
where the $O(1)$ initial contribution vanishes after division by $Nh$ and taking $N\to\infty$.

\subsubsection{Numerical Results}
To parameterize the deviation of $R$ from the identity by $p_0$, we set all diagonal entries to $1-p_0$ and all off-diagonal entries to $p_0/2$.

\begin{figure}[!bth]
    \centering
    \includegraphics[width=\columnwidth]{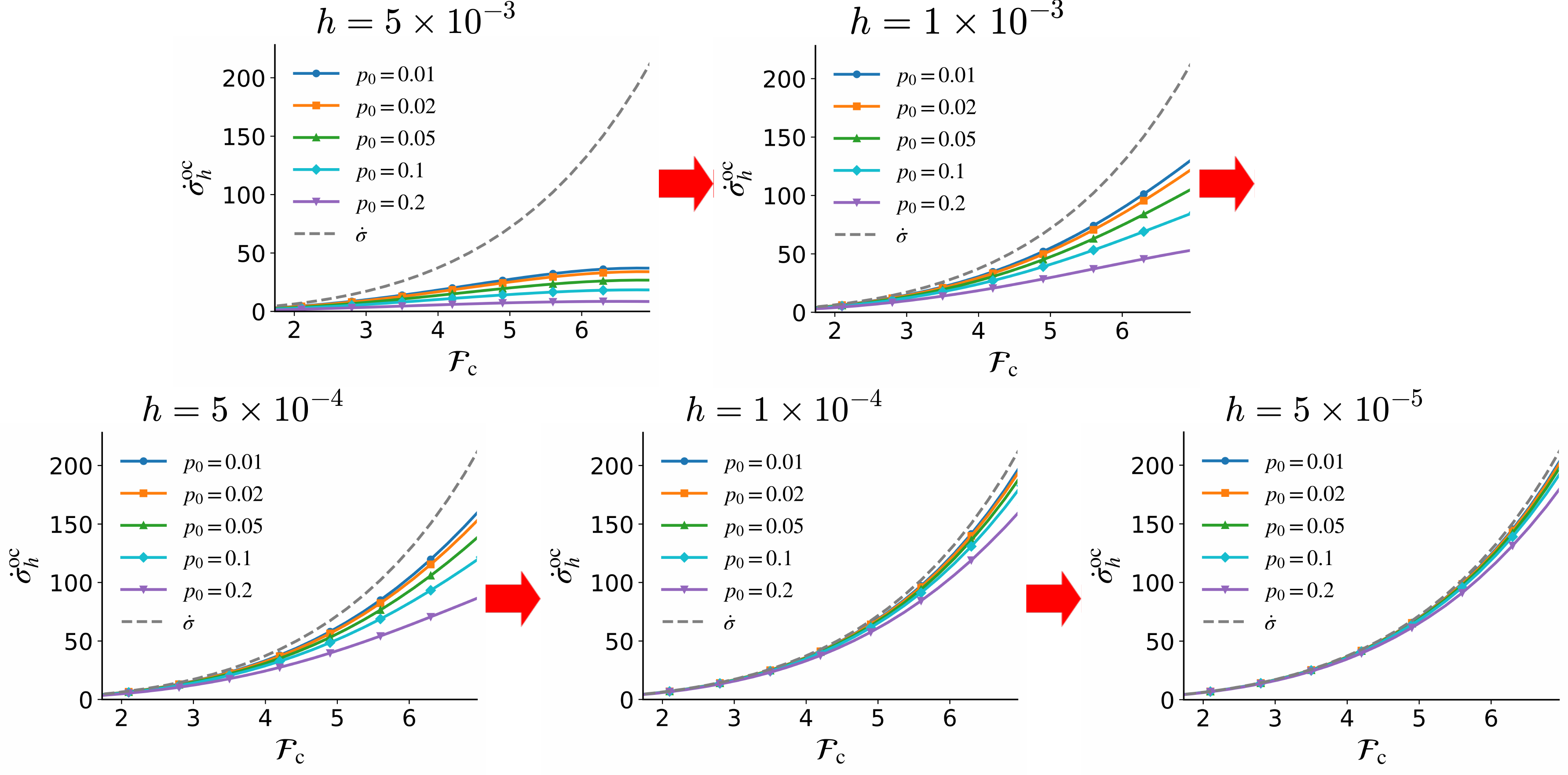}
    \caption{Effect of the recoloring matrix $R$ and sampling interval $h$ on $\dot \sigma^\mathrm{oc-appro}_h$ under different cycle affinities. The settings are shown in the subsection ``4. Description of Numerical Simulation Setting and Approximation''. To avoid verbosity in the figure, ``-appro'' is omitted}
    \label{h0}
\end{figure}

For \(h=0.005, 0.001, 0.0005, 0.0001,\) and \(0.00005\), 
we plotted the theoretical values of $\dot{\sigma}^{\mathrm{oc-appro}}_h$ as functions of $\mathcal{F}_{\mathrm c}$ for several values of $p_0$, without simulating stochastic trajectories (see Fig. \ref{h0}).
The dashed line represents $\dot{\sigma}$. 
In Fig. \ref{h0}, as $h$ decreases, each $\dot{\sigma}^{\mathrm{oc-appro}}_h$–$\mathcal{F}_{\mathrm c}$ curve increasingly matches the $\dot{\sigma}$–$\mathcal{F}_{\mathrm c}$ curve in both shape and position.
This numerical trend is consistent with Eq. \eqref{dense-limit-from-h}: $\dot\sigma_h^{\mathrm{oc\text{-}appro}}\to\dot\sigma$ as $h\to0$.
Moreover, under the stated conditions on $R$, $\dot{\sigma}^{\mathrm{oc}}$ coincides with $\dot{\sigma}$, consistent with Eq. (12) in the main text.

Equations \eqref{dense-limit-from-h} and (12) in the main text imply that these curves eventually converge to $\dot{\sigma}$ for all \(p_0\neq 2/3\).
With this in mind, we can examine more properties from the simulation results, other than Eq. \eqref{dense-limit-from-h} and Eq. (12).
Observing the curves for different values of $p_0$ in the same figure, the curves corresponding to smaller values of $p_0$ are also closer to the dashed $\dot{\sigma}$ curve, whereas even a modest increase in $p_0$ substantially enlarges the deviation from it. These results therefore suggest that convergence becomes more difficult as the color–-channel ambiguity increases. Consequently, using a condition whose state-to-observable mapping does not deviate substantially from one-to-one may enable efficient thermodynamic inference at relatively low experimental and computational cost, supporting our tentative recommendation in the main text.

\subsubsection{Description of Numerical Simulation Setting and Approximation}
Numerically, the invariant measure $\mu_h$ is represented by a finite weighted collection of forward--reverse filter pairs,
\begin{equation}
    \mu_h
    \simeq
    \sum_{a=1}^{M}
    w_a\,
    \delta_{(\alpha_a,\beta_a)},
    \qquad
    \sum_{a=1}^{M}w_a=1.
    \label{finite-filter-representation}
\end{equation}
Here, $a$ labels a deterministic observation branch, or equivalently a filter particle. The vectors
\begin{subequations}
\begin{align}
    \alpha_{a,i}
    &=
    \Pr(X_{kh}=i\mid J_{0:k})
    ,
    \\
    \beta_{a,i}
    &=
    \Pr^\dagger(X_{kh}=i\mid J_{0:k})
\end{align}
\end{subequations}
are, respectively, the posterior probabilities of the microscopic state under the forward and reversed models, conditioned on the same observed label history represented by particle $a$. The weight $w_a$ is the total forward probability of the observation histories represented by that particle. Forward weights are used because $D_{\mathrm{KL}}(\mathbb P\Vert\mathbb P^\dagger)$ is averaged with respect to the forward process.

For the three-state model,
\begin{equation}
    \alpha_a
    =
    (\alpha_{a,1},\alpha_{a,2},\alpha_{a,3}),
    \qquad
    \beta_a
    =
    (\beta_{a,1},\beta_{a,2},\beta_{a,3}).
\end{equation}
Since both filters are normalized,
\begin{equation}
    \alpha_{a,3}
    =
    1-\alpha_{a,1}-\alpha_{a,2},
    \qquad
    \beta_{a,3}
    =
    1-\beta_{a,1}-\beta_{a,2}.
\end{equation}
Each filter pair can therefore be represented by the four-dimensional
coordinate
\begin{equation}
    z_a
    =
    (\alpha_{a,1},\alpha_{a,2},
     \beta_{a,1},\beta_{a,2}).
\end{equation}

Deterministic branching over all possible next labels causes the number of
filter particles to increase rapidly. To keep the representation finite,
the four-dimensional filter space is divided into grid cells, hereafter
called bins. The grid coordinate of particle $a$ is
\begin{equation}
    b(z_a)
    =
    \operatorname{round}
    \left(
        \frac{z_a}{\delta_{p_0}}
    \right),
    \label{filter-bin-coordinate}
\end{equation}
where the rounding is applied componentwise and $\delta_{p_0}$ is the grid
spacing. A bin with grid coordinate $q$ is the set
\begin{equation}
    B_q
    =
    \left\{
        a:
        b(z_a)=q
    \right\}.
\end{equation}
Thus, two filter particles are assigned to the same bin when all four of
their rounded coordinates coincide.

For each nonempty bin $B$, the total probability weight is
\begin{equation}
    w_B
    =
    \sum_{a\in B}w_a.
\end{equation}
All filter particles in that bin are then replaced by one representative
filter pair, defined by the probability-weighted posterior averages
\begin{subequations}
\begin{align}
    \alpha_B
    &=
    \frac{1}{w_B}
    \sum_{a\in B}w_a\alpha_a,
    \\
    \beta_B
    &=
    \frac{1}{w_B}
    \sum_{a\in B}w_a\beta_a.
\end{align}
\label{weighted-filter-merging}
\end{subequations}
The merged particle $(\alpha_B,\beta_B)$ therefore carries the total
weight $w_B$. This binning procedure is a numerical approximation used to
represent the filter distribution; it is not an additional physical
coarse-graining of the microscopic or observed dynamics.

For the figure of $h=0.005$, the filter grid spacing used in the calculation is
\begin{equation}
    \delta_{p_0} = 0.0025.
\end{equation}
However, for \(h=0.001, 0.0005, 0.0001,\) and \(0.00005\), the setting above is so computationally expensive that it is better to introduce self-adaptive mechanism. As a result, the filter grid spacings used in the calculation are
\begin{equation}
\begin{split}
    \delta_{0.01}&=0.0025,
    \qquad
    \delta_{0.02}=0.005,
    \qquad
    \delta_{0.05}=0.01,
    \\
    \delta_{0.1}&=0.02,
    \qquad
    \delta_{0.2}=0.04.
\end{split}
\label{filter-grid-spacings}
\end{equation}

\subsection{Settings of Fig.S2: Parameters}
We employ the following recoloring matrix for the reconstruction:
\begin{equation}
    R_0 = \begin{pmatrix} 0.99 & 0.005 & 0.01\\ 0.005 & 0.99 & 0.01\\ 0.005 & 0.005 & 0.98\end{pmatrix}.
\end{equation}

When calculating the estimators, we set the sampling time step as 0.005.

\subsection{Settings of Fig.2-(b) in the Main Text: Optimization of $\Delta t_{\mathrm{opt}-n}$ and $(\{q_k\}_{k=0}^n)_{\mathrm{opt}}$}
To obtain the maximal reconstruction at each order $n$, we determine $\Delta t_{\mathrm{opt}-n}$ and $(\{q_k\}_{k=0}^n)_{\mathrm{opt}}$. Following the methodology in End Matter, these quantities are obtained via numerical exploration of the whole full parameter space, whose cross-section is shown in Fig. \ref{opt}. The resulting $\Delta t_{\mathrm{opt}-n}$ and $(\{q_k\}_{k=0}^n)_{\mathrm{opt}}$ for different $K$ are listed in Table \ref{optimization_data}. We emphasize that $\Delta t_{\mathrm{opt}-n}$ and $(\{q_k\}_{k=0}^n)_{\mathrm{opt}}$ depend not only on $n$, but also on $K$ and $R_0$.

\begin{table}
    \centering
    \begin{tabular}{|c|c|c|c|}
    \hline
     $k^+/k^-$ & $n$ & $\Delta t_{\mathrm{opt}-n}$ & $(\{q_k\}_{k=0}^n)_{\mathrm{opt}}$ \\
    \hline
     \multirow{4}{*}{2} & 1 & 0.003260 & \{0, 1\} \\
    \cline{2-4}
      & 2 & 0.003700 & \{0, 0.511429, 1\} \\
    \cline{2-4}
      & 3 & 0.000659 & \{0, 0.006572, 0.996840, 1\} \\
    \cline{2-4}
      & 4 & 0.000661 & \{0, 0.002781, 0.516312, 0.997219, 1\} \\
    \hline
     \multirow{4}{*}{3} & 1 & 0.002565 & \{0, 1\} \\
    \cline{2-4}
      & 2 & 0.003466 & \{0, 0.497366, 1\} \\
    \cline{2-4}
      & 3 & 0.000329 & \{0, 0.001548, 0.994497, 1\} \\
    \cline{2-4}
      & 4 & 0.000311 & \{0, 0.002751, 0.512347, 0.997249, 1\} \\
    \hline
     \multirow{4}{*}{6} & 1 & 0.001508 & \{0, 1\} \\
    \cline{2-4}
      & 2 & 0.001978 & \{0, 0.498304, 1\} \\
    \cline{2-4}
      & 3 & 0.000116 & \{0, 0.003697, 0.998960, 1\} \\
    \cline{2-4}
      & 4 & 0.000112 & \{0, 0.001946, 0.544414, 0.996947, 1\} \\
    \hline
     \multirow{4}{*}{9} & 1 & 0.000988 & \{0, 1\} \\
    \cline{2-4}
      & 2 & 0.001462 & \{0, 0.497947, 1\} \\
    \cline{2-4}
      & 3 & 0.000117 & \{0, 0.000571, 0.996301, 1\} \\
    \cline{2-4}
      & 4 & 0.000170 & \{0, 0.002262, 0.567930, 0.997738, 1\} \\
    \hline
    \end{tabular}
    \caption{Numerical optimization results. The exploration algorithm over $\Delta t$ and $\{q_k\}$ was performed using an adaptive search. The total lag $\Delta t$ was first scanned on a mixed logarithmic-linear grid, while the relative intervals $\{q_k\}$ were searched on an unbiased coarse simplex grid and then adaptively refined around high-scoring candidates. A final local pattern search was applied in $\log \Delta t$ and simplex coordinates. The reported optimum is the best value found under this adaptive search resolution and evaluation budget, rather than a certified global optimum of the continuous parameter space. All reported values are direct numerical outputs (subject to numerical precision limits), including the border values ($q_i-q_{i-1} < 0.005$) which may result in a zero sampling time step when calculating estimators.}
    \label{optimization_data}
\end{table}

Using these optimized parameters, we can explain why the optimized reconstructions of $\dot\sigma^{\mathrm{est-}3}_\mathrm{opt}$ and $\dot\sigma^{\mathrm{est-}4}_\mathrm{opt}$ in the main text are large-variance. For $n=3,4$, boundary values of $q_k$ are approached ($q_0 \simeq q_1$ and $q_{n-1} \simeq q_n$), rendering the reconstructions highly sensitive to short-time fluctuations in stochastic trajectories, which vary substantially across realizations.

Caution: $\Delta t$ and $\{q_k\}_{k=0}^n$ are interdependent during optimization; thus, sequential optimization is invalid. A simple example is as follows: when $k^+/k^-=2$, we find $\dot\sigma^{\mathrm{est-}3}_{0.0072, \{0, \frac{1}{3}, \frac{2}{3}, 1\}}(t) > \dot\sigma^{\mathrm{est-}3}_{0.0072, \{0, 0, 1, 1\}}(t)$, whereas $\dot\sigma^{\mathrm{est-}3}_{0.0024, \{0, \frac{1}{3}, \frac{2}{3}, 1\}}(t) < \dot\sigma^{\mathrm{est-}3}_{0.0024, \{0, 0, 1, 1\}}(t)$.

\subsection{Experimental Feasibility for Fluorescence Correlation Spectroscopy}
Single-molecule fluorescence correlation spectroscopy (FCS) could be used for observation in this setting \cite{qian2004fluorescence, qian2009chemical, haustein2007fluorescence, bacia2006fluorescence}, allowing us to map microscopic state transitions onto observables. The reliability and practical utility of fluorescence correlation spectroscopy rest on two key assumptions:
\begin{enumerate}
    \item As in previous single-molecule studies \cite{qian2004fluorescence}, to simplify the analysis, we assume the single-molecule fluorescence correlation spectroscopy signal originates exclusively from microstate transitions. Diffusive contributions are eliminated by positional fixation, achievable in specialized setups \cite{widengren2006single, krichevsky2002fluorescence}.
    \item Following Ref. \cite{song2024information}, we reduce photophysical dynamics to effective Poissonian emission, yielding a minimal yet sufficient description. Photon emission is assumed to follow independent Poisson statistics, irrespective of color, and is presumed not to perturb the underlying state transitions.
\end{enumerate}

This experiment may also be implemented using other single-molecule fluorescence measurements \cite{talaga2007markov}.

\section{Advantage over the Pseudo-EPR}
\subsection{Calculating $\lim_{\Delta t \to 0} \dot\sigma^{\mathrm{est-}n}_{\Delta t, \{q_k\}_{k=0}^{n}}(t)$}
For the 2-order reconstruction, we find
\begin{equation}
    D_{\{J_0, J_1\}} = \sum_{u, v} \mathcal{O}_{u}^{J_0} \mathcal{O}_{v}^{J_1} K_{vu} p_u^{\mathrm{ss}} = \sum_{i',i} p_{J'\gets i'}^\mathrm{map}p_{J\gets i}^\mathrm{map}K_{i'i}p_i(t) = \mathcal{J}_{J_1J_0},
\end{equation}
which can be interpreted as the directed flux from $J_0$ to $J_1$.
As a result, according to Eq. \eqref{prob_expansion}, when $S_{\{J_0, J_1\}}=0$ holds for all pairs of $\{J_0, J_1\}$ where $J_0 \ne J_1$ (i.e. when the state--observable relation \textbf{is one-to-one} or corresponds to \textbf{deterministic lumping}) in steady states, it turns out that
\begin{equation}
    \begin{split}
        \dot\sigma^{\mathrm{est-}1}_{\Delta t, \{0,1\}}(t)\cdot\Delta t &= \sum_{\omega} P(\omega) \ln \frac{P(\omega)}{P(\omega^\dagger)} \\
        &= \sum_{J_0 \ne J_1} \left[ D_{\{J_0, J_1\}} \Delta t + \mathrm{o}(\Delta t) \right] \ln \frac{D_{\{J_0, J_1\}} \Delta t + \mathrm{o}(\Delta t)}{D_{\{J_1, J_0\}} \Delta t + \mathrm{o}(\Delta t)}\\
        &= \sum_{J_0\ne J_1} \left[ \mathcal{J}_{J_1 J_0} \Delta t + \mathrm{o}(\Delta t) \right] \ln \frac{\mathcal{J}_{J_1 J_0}+ \mathrm{o}(1)}{\mathcal{J}_{J_0 J_1}+ \mathrm{o}(1)},
    \end{split}
\end{equation}
where the second equality holds because the pairs $\{J_0, J_1\}$ with $J_0 = J_1$ give zero-value terms according to Eq. \eqref{s0}. Therefore,
\begin{equation}
    \lim_{\Delta t \to 0} \dot\sigma^{\mathrm{est-}1}_{\Delta t, \{0,1\}}(t) = \lim_{\Delta t \to 0} \frac{\dot\sigma^{\mathrm{est-}1}_{\Delta t, \{0,1\}}(t)\cdot\Delta t}{\Delta t} = \sum_{J_0\ne J_1} \mathcal{J}_{J_1 J_0} \ln \frac{\mathcal{J}_{J_1 J_0}}{\mathcal{J}_{J_0 J_1}} 
    \label{lim=oc}
\end{equation}
Furthermore, when $n \ge 1$, we obtain $\dot\sigma^{\mathrm{est-}n}_{\Delta t, \{q_k\}_{k=0}^n}(t) \ge \dot\sigma^{\mathrm{est-}1}_{\Delta t, \{0,1\}}(t)$ from Eq. \eqref{plug-in}. So we finally get
\begin{equation}
    \lim_{\Delta t \to 0} \dot\sigma^{\mathrm{est-}n}_{\Delta t, \{q_k\}_{k=0}^n}(t) \ge \sum_{J_0\ne J_1} \mathcal{J}_{J_1 J_0} \ln \frac{\mathcal{J}_{J_1 J_0}}{\mathcal{J}_{J_0 J_1}}.
\end{equation}

We should emphasize here that Eq. \eqref{lim=oc} is violated when $S_{\{J_0, J_1\}}=0$ \textbf{no longer holds}.

\subsection{Advantage over pseudo-EPR with One-to-one State--observable Relation}
According to Ref. \cite{shiraishi2021optimal}, the pseudo-EPR is defined as
\begin{equation}
    \dot\sigma^\mathrm{pseudo}(t) = \sum_{i \ne i'} \dot\Pi_{ii'} = \sum_{i \ne i'} \frac{(K_{ii'}p_{i'}(t)-K_{i'i}p_{i}(t))^2}{K_{ii'}p_{i'}(t)+K_{i'i}p_{i}(t)} = \sum_{i \ne i'} \frac{(\mathcal{J}_{ii'}-\mathcal{J}_{i'i})^2}{\mathcal{J}_{ii'}+\mathcal{J}_{i'i}}.
\end{equation}
Thus, when the state--observable relation is one-to-one, we obtain
\begin{equation}
    \lim_{\Delta t \to 0} \dot\sigma^{\mathrm{est-}n}_{\Delta t, \{q_k\}_{k=0}^n}(t) \ge \sum_{J_0 \ne J_1} \mathcal{J}_{J_1 J_0} \ln \frac{\mathcal{J}_{J_1 J_0}}{\mathcal{J}_{J_0 J_1}} = \sum_{i \ne i'} \mathcal{J}_{i'i} \ln \frac{\mathcal{J}_{i' i}}{\mathcal{J}_{ii'}} \ge \sum_{i \ne i'} \frac{(\mathcal{J}_{ii'}-\mathcal{J}_{i'i})^2}{\mathcal{J}_{ii'}+\mathcal{J}_{i'i}} = \dot\sigma^\mathrm{pseudo}(t),
    \label{lim-pseudo}
\end{equation}
where $J_0 \to i$ and $J_1 \to i'$ are based on the one-to-one state--observable relation, and the inequality follows from $a\ln\frac{a}{b} + b\ln\frac{b}{a} \ge \frac{2(a-b)^2}{a+b}$, which holds for all $a, b >0$. This is consistent with Eq. \eqref{eq:S-oc-equals} and the well-known fact that $\dot\sigma_\mathrm{pseudo}(t) \le \dot\sigma(t)$ \cite{shiraishi2021optimal}.

%We should mention here that, the former works, which operate inference from irreversibility measure \cite{roldan2010estimating, roldan2021quantifying, kapustin2024utilizing}, also share this advantage.

\subsection{Why the Advantage Holds Far from Equilibrium}
We now explain the statement in the main text that the advantage is particularly pronounced under conditions far from equilibrium.

We define the thermodynamic force
\begin{equation}
    \mathcal F_{ii'} = \ln\frac{\mathcal{J}_{ii'}}{\mathcal{J}_{i'i}}
\end{equation}
for each microscopic transition path, such that 
\begin{equation}
    \begin{split}
        &\dot\Pi_{ii'} = (\mathcal{J}_{ii'}+\mathcal{J}_{i'i}) \tanh^2 \frac{\mathcal F_{ii'}}{2},\\
        &\dot\sigma_{ii'} = (\mathcal{J}_{ii'}+\mathcal{J}_{i'i}) \frac{\mathcal F_{ii'}} 2 \tanh \frac{\mathcal F_{ii'}}{2}.
    \end{split}
\end{equation}
Therefore, 
\begin{equation}
    \frac{\dot\sigma_{ii'}}{\dot\Pi_{ii'}} = \frac{\mathcal F_{ii'}/2}{\tanh (\mathcal F_{ii'}/2)} = \frac{\left | \mathcal F_{ii'} /2\right | }{\tanh ( \left | \mathcal F_{ii'} \right | /2)}.
\end{equation}
Denoting $\mathcal F^* = \min\{\left | \mathcal F_{ii'} \right |\}$ and $c^* = \mathcal F^*/(2\tanh (\mathcal F^*/2))$ ($c^*\coloneqq1$ when $F^*=0$), we obtain
\begin{equation}
    \dot\sigma(t) = \sum_{i \ne i'}\dot\sigma_{ii'} \ge \sum_{i \ne i'} c^*\dot\Pi_{ii'} = c^*\dot\sigma^\mathrm{pseudo}(t),
    \label{ge-pseudo}
\end{equation}
which is guaranteed by the strictly increasing property of $f(x) = x/\tanh (x/2)$ on the positive half-axis.

For a one-to-one mapping, Eqs. \eqref{14} and \eqref{lim-pseudo} give $\lim_{\Delta t\to0}\dot\sigma^{\rm est-n}=\dot\sigma$, so Eq. \eqref{ge-pseudo} applies directly to this limit; for other mappings, Eq. \eqref{ge-pseudo} alone does not bound the estimator.

\subsection{How Experimental Ambiguity also Suppresses Inference Based on Thermodynamic Uncertainty Relation}
According to Ref. \cite{gingrich2016dissipation}, when constructing a TUR through a generalized current, to make the TUR tighter, one should choose the weight $d(i,i')$ of transition path as close as possible to being proportional to the thermodynamic force:
\begin{equation}
    \mathcal{J}^\mathrm{general} = \sum _{i\ne i'} d(i,i') \mathcal{J}_{ii'}, \qquad d(i,i')_\mathrm{opt} \propto \mathcal F_{ii'}.
\end{equation}

Partial observations can restrict the accessible microscopic currents. For a fixed class of valid current-based TURs, let $B(d)$ denote the bound and $\mathcal D_{\mathrm{obs}}\subseteq\mathcal D_{\mathrm{full}}$ the accessible and unrestricted weight sets. Then
\begin{equation*}
    \sup_{d\in\mathcal D_{\mathrm{obs}}}B(d)
    \le\sup_{d\in\mathcal D_{\mathrm{full}}}B(d).
\end{equation*}
Thus ambiguity can weaken optimized current-based inference. For random-mapping readouts, a TUR must be justified for the actual measured current, since label changes need not be microscopic jumps. This supports the possibility discussed in the main text.

\bibliography{refs}